\documentclass[fleqn,usenatbib]{mnras}

\usepackage{newtxtext,newtxmath}
\usepackage{booktabs}
\usepackage{physics}
\usepackage{braket}
\usepackage{soul}

\usepackage[T1]{fontenc}

\DeclareRobustCommand{\VAN}[3]{#2}
\let\VANthebibliography\thebibliography
\def\thebibliography{\DeclareRobustCommand{\VAN}[3]{##3}\VANthebibliography}

\newcommand{\comment}[1]{}

\usepackage[most]{tcolorbox}
\newtcolorbox{rbox}[1]{colback=red!5!white,colframe=red!75!black,fonttitle=\bfseries,title=#1}
\newtcolorbox{gbox}[1]{colback=gray!5!white,colframe=gray!75!black,fonttitle=\bfseries,title=#1}

\usepackage{graphicx}	
\usepackage{amsmath}	
\usepackage{multirow}
\usepackage{caption}
\usepackage{academicons}
\definecolor{orcidlogocol}{HTML}{A6CE39}
\usepackage{cancel}

\newcommand{\galaxyrate}{{\sc galaxy$\mathcal{R}$ate}}
\newcommand{\cosmorate}{{\sc cosmo$\mathcal{R}$ate}}

\title[Host galaxies of binary compact objects]{Modelling the host galaxies of binary compact object mergers with observational scaling relations}

\author[F. Santoliquido et al.]{Filippo Santoliquido,$^{1,2}$\thanks{E-mail: filippo.santoliquido@phd.unipd.it}
Michela Mapelli,$^{1,2,3}$\thanks{E-mail: michela.mapelli@unipd.it}
M. Celeste Artale,$^{1,2,4,5}$ and
Lumen Boco$^{6,7,8}$
\\
$^{1}$Physics and Astronomy Department Galileo Galilei, University of Padova, Vicolo dell'Osservatorio 3, I--35122, Padova, Italy\\
$^{2}$INFN--Padova, Via Marzolo 8, I--35131 Padova, Italy\\
$^{3}$INAF--Osservatorio Astronomico di Padova, Vicolo dell'Osservatorio 5, I--35122, Padova, Italy\\
$^{4}$Institut f{\"u}r  Astro- und Teilchenphysik, Universit{\"a}t Innsbruck, Technikerstrasse 25/8, A-6020, Innsbruck, {\"O}sterreich\\
$^{5}$Department of Physics and Astronomy, Purdue University, 525 Northwestern Avenue, West Lafayette, IN 47907, USA \\
$^{6}$SISSA, Via Bonomea 265, I-34136 Trieste, Italy\\
$^{7}$IFPU - Institute for fundamental physics of the Universe, Via Beirut 2, I-34014 Trieste, Italy\\
$^{8}$INFN-Trieste, via Valerio 2, I-34127 Trieste, Italy\\
}

\date{Accepted XXX. Received YYY; in original form ZZZ}

\pubyear{2021}

\begin{document}
\label{firstpage}
\pagerange{\pageref{firstpage}--\pageref{lastpage}}
\maketitle

\begin{abstract}

\noindent The merger rate density evolution of binary compact objects  and the properties of their host galaxies carry crucial information to understand the sources of gravitational waves. Here, we present \galaxyrate{}, a new code that estimates the merger rate density of binary compact objects and the properties of their host galaxies, based on observational scaling relations. We generate our synthetic galaxies according to the galaxy stellar mass function. 
We estimate the metallicity according to both the mass-metallicity relation (MZR) and the fundamental metallicity relation (FMR). Also, we take into account galaxy-galaxy mergers and the evolution of the galaxy properties from the formation to the merger of the binary compact object. 
We find that the merger rate density changes dramatically depending on the choice of the star-forming galaxy main sequence, especially in the case of binary black holes (BBHs) and black hole neutron star systems (BHNSs). The slope of the merger rate density of BBHs and BHNSs is steeper if we assume the MZR with respect to the FMR, because the latter predicts a shallower decrease of metallicity with redshift. In contrast, binary neutron stars (BNSs) are only mildly affected by both the galaxy main sequence and  metallicity relation. Overall,  BBHs and BHNSs tend to form in low-mass metal-poor galaxies and merge in high-mass metal-rich galaxies, while BNSs form and merge in massive galaxies. We predict that passive galaxies host at least $\sim{5-10}$\%, $\sim{15-25}$\%, and $\sim{15-35}$\% of all BNS, BHNS and BBH mergers in the local Universe.

\end{abstract}




\section{Introduction}

The third gravitational wave transient catalog (GWTC-3) of the LIGO--Virgo--KAGRA collaboration (LVK)  contains 90 gravitational-wave (GW) event candidates \citep{O3b}. 
From this growing number of detections, we can extract several astrophysical properties of binary compact objects (BCOs), such as their masses, spins and merger rates. 
From GWTC-3, the LVK  inferred a local merger rate density  $\mathcal{R}_0^{\rm{BBH}} = 16 - 61$ Gpc$^{-3}$ yr$^{-1}$, $\mathcal{R}_0^{\rm{BHNS}} = 7.8 - 140$ Gpc$^{-3}$ yr$^{-1}$ and $\mathcal{R}_0^{\rm{BNS}} = 10-1700$ Gpc$^{-3}$ yr$^{-1}$ for binary black holes (BBHs), black hole-neutron star binaries (BHNSs) and binary neutron stars (BNSs), respectively \citep{pop3_2021}. 
In the case of BBHs, it is even possible to reconstruct the evolution of the merger rate with redshift $\mathcal{R}(z) \propto{} (1+z)^k $ with $z \leq{} 1$.  \cite{pop3_2021} find $k > 0$ at 99.6\% credibility, indicating that the BBH merger rate density increases with redshift. 
Thanks to  the next-generation ground-based GW detectors, Einstein Telescope \citep{punturo2010} and Cosmic Explorer \citep{reitze2019}, we will be able to reconstruct the redshift evolution of the BBH merger rate  up to redshift $z=10$ or even higher \citep{kalogera2019, kalogera2021, maggiore2020}. 

While the merger rate provides crucial insights about the formation of BCOs \citep[e.g.,][]{dominik2013,belczynski2016,mapelli2017,mapelli2018,baibhav2019,neijssel2019}, the properties of their host galaxies (HGs) represent another fundamental piece of information \citep[e.g.,][]{perna2002,belczynski2006,oshaughnessy2010,lamberts2016,schneider2017,mapelli2018b,artale2019,artale2020a,artale2020b}. Also, the identification of  
the host galaxy is crucial 
to reduce the uncertainties on the measure of the Hubble constant from GW sources \citep{h0_gw170817, Fishbach2019, gray2020, Jin2021, Hebertt2022}, 
and to successfully use BCOs as tracers of large scale structures \citep{vijaykumar2020, susmita2020, Libanore2021,Libanore2022,Mukherjee2021, Cigarr2022}. 

At present, only the HG of the BNS merger GW170817 \citep{bns2017,abbott2017e, Goldstein2017,Savchenko2017, margutti2017, Coulter2017, Soares-Santos2017, Chornock2017, Cowperthwaite2017,Nicholl2017, Pian2017, Alexander2017}, the elliptical galaxy NGC~4993, has been identified beyond any reasonable doubt \citep{levan2017, Im2017, Ebrova2020, Kilpatrick2022}. 
The main obstacle to the successful identification of the HG is represented by the sky localisation uncertainties of GW detectors, currently being of the order of several ten square degrees \citep{abbott201prospects}.

Several criteria to rank the galaxies within the sky localisation region have been proposed, in order to  optimize the search for  electromagnetic counterparts 
\citep[e.g.,][]{Kopparapu2008, arcavi2017, ducoin2020, Stachie2020,artale2020b,Ashkar2021b, Ashlar2021a, Kovlakas2021, perna2021}. 
Furthermore, several authors studied 
the properties of the HGs on a 
theoretical ground;  most of them interface the outputs of cosmological simulations with catalogs of compact objects, either obtained through population-synthesis or phenomenological models \citep[e.g.,][]{Shaughnessy2017, schneider2017,mapelli2017,toffano2019, artale2019,  artale2020a, artale2020b, susmita2020, mandhai2021, chu2021, rose2021,perna2021,Mukherjee2021b}. These previous works suggest that the BCO merger rate per galaxy  correlates with the stellar mass and star formation rate (SFR) of the HG  \citep[e.g.,][]{artale2020a}.

Here, we present a new fast numerical tool to estimate the cosmic merger rate density and to characterize the properties of the HGs of BCO mergers. Our new code \galaxyrate{}  exploits the main observational scaling relations (galaxy stellar mass function, SFR distribution, mass-metallicity relation, and fundamental-metallicity relation) to generate the distribution of galaxy masses, SFR and metallicity across cosmic time \citep{Boco2019, Safarzadeh2019, Safarzadeh2019a, Elbert2018,chruslinska2019, chruslinska2020, chruslinska2021}. We derive the properties of the BCOs, and especially their delay times\footnote{We define the delay time as the time elapsed from the formation of the progenitor binary star to the merger of the BCO.} $t_{\rm del}$, from up-to-date binary population synthesis simulations \citep{santoliquido2021}. 
\galaxyrate{} is very flexible and can read catalogs from phenomenological models of BCO mergers as well. 

Unlike models based on computationally expensive cosmological simulations, \galaxyrate{} can be used to probe the parameter space of BCO mergers: a single model requires $\sim 15$ CPU hours on a single core. With respect to similar codes based on observational scaling relations \citep[e.g.,][]{Boco2019, Safarzadeh2019, Safarzadeh2019a, Elbert2018}, we have developed a new algorithm that is able to differentiate between the galaxy in which a BCO forms [hereafter, formation galaxy (FG)], and the galaxy where the BCO merges after the delay time [hereafter, host galaxy (HG)]. In fact, the properties of the FG and those of the HG can be very different from each other, not only because galaxies merge with 
other galaxies across the cosmic time, but also because the same galaxy can evolve significantly during the BCO delay time, changing its mass, SFR and metallicity. Our new algorithm evaluates a conditional probability that the HG of a BCO has mass $M_{\rm{host}}$ and star formation rate  ${\rm{SFR}}_{\rm{host}}$,  given the mass of the FG  ($M_{\rm{form}}$), its SFR (${\rm SFR}_{\rm{form}}$), the formation and the merger redshift of the BCO ($z_{\rm form}$ and $z_{\rm merg}$, respectively). To calculate this probability, we use the galaxy merger tree extracted from the {\sc eagle} cosmological simulation with a (100~cMpc)$^3$ volume \citep{schaye2015,Qu2017}, but our formalism can be easily generalized to other merger trees.

The paper is organized as follows: in Section~\ref{sec:pop-synth}, we present the main formalism we adopted in our population-synthesis code {\sc{mobse}} \citep{giacobbo2018b}; Section~\ref{sec:galaxymodel} describes the observational scaling relations adopted in \galaxyrate{}; Sections~\ref{sec:MRD} and  \ref{sec:mergergalaxies} discuss the method to evaluate the merger rate density and the conditional probability, respectively. Section~\ref{sec:results} presents our main results. We discuss the implications of our new methodology in Section~\ref{sec:discussion}, and draw our main conclusions in Section~\ref{sec:conlusion}.

\section{Methods}
\label{sec:methods} 

Here, we describe our new code \galaxyrate{}, which calculates the merger rate evolution of BCOs and the properties of their host galaxies, based on population-synthesis simulations and observational scaling relations. \galaxyrate{} is an upgrade of our code \cosmorate{} \citep[][hereafter S20]{santoliquido2020}. Appendix \ref{sec:cosmorate} is a short description of the main features of \cosmorate{}. Table \ref{tab:param} summarises the parameters included in \galaxyrate{} and discussed in the following sections.  
 
\subsection{Binary compact objects (BCOs)}
\label{sec:pop-synth}

The open-source population synthesis code {\sc mobse} \citep{mapelli2017,giacobbo2018a} is an upgraded and customized version of {\sc bse} \citep{hurley2000,hurley2002}. With respect to the original version of {\sc bse}, {\sc mobse} includes an up-to-date formalism for stellar winds \citep{giacobbo2018a}, and several new models for the outcome of electron-capture \citep{giacobbo2019}, core-collapse \citep{fryer2012} and pair-instability supernovae \citep{spera2017,mapelli2020}. Here, we assume the rapid core-collapse supernova model \citep{fryer2012}, which enforces a mass gap between the maximum mass of a neutron star (2 M$_\odot$) and the minimum mass of a black hole (5 M$_\odot$).  Finally, we model natal kicks of compact remnants (both neutron stars and black holes) as $v_{\rm kick}\propto{}m_{\rm ej}/m_{\rm rem}\,{}v_{\rm H05}$, where $m_{\rm rem}$ is the mass of the compact object, $m_{\rm ej}$ is the mass of the ejecta, and $v_{\rm H05}$ is a random number following a Maxwellian distribution with one-dimensional root-mean square $\sigma_{\rm H05}=265$ km s$^{-1}$ \citep{giacobbo2020}, inspired from the proper motions of Galactic young pulsars \citep{hobbs2005}. {\sc mobse} itegrates the main binary evolution processes (wind mass transfer, Roche lobe overflow, common envelope, magnetic braketing, equilibrium tides and GW decay) as described in \cite{hurley2002}. We refer to \cite{giacobbo2018a} for further details on {\sc mobse}.

For the runs presented in this paper, we have used three different models, with three different values of the common-envelope parameter $\alpha=1,$ 3 and 5 (hereafter, $\alpha{}1$, $\alpha{}3$ and $\alpha{}5$). For each run, we have simulated 12 different metallicities: $Z=0.0002$, 0.0004, 0.0008, 0.0012, 0.0016, 0.002, 0.004, 0.006, 0.008, 0.012, 0.016, and 0.02. For metallicities $Z\leq{}0.002$ ($Z>0.002$), we have simulated $10^7$ ($2\times{}10^7$) massive binary stars per each model. 

The primary mass of each binary star follows a Kroupa initial-mass function \citep[IMF,][]{kroupa2001} with minimum (maximum) mass $=5$ (150) M$_\odot$.  We derive the mass ratio $q=m_2/m_1$ as $\mathcal{F}(q) \propto q^{-0.1}$ with $q\in [0.1,\,{}1]$, the orbital period $P$ from $\mathcal{F}(\Pi) \propto \Pi^{-0.55}$ with $\Pi = \log_{10}(P/\text{day}) \in [0.15\,{}5.5]$ and the eccentricity $e$ from $\mathcal{F}(e) \propto e^{-0.42}~~\text{with}~~ 0\leq e \leq 0.9$ \citep{sana2012}. Appendix~\ref{sec:delay} discusses some of the main features of these population-synthesis simulations (merger  efficiency and delay time distribution).

\subsection{Observational scaling relations}
\label{sec:galaxymodel}

Hereafter, we call formation galaxy (FG) the galaxy where the 
stellar progenitors of the BCO form, and host galaxy  (HG) the galaxy where the BCO merges via GW emission. 
We describe each FG with three parameters: 
the galaxy stellar mass ($M_*$), star-formation rate (SFR) and a log-normal  distribution of metallicities. This is the minimum amount of information that allows us to understand the link between galaxy properties and  compact object mergers. 
In order to create a population of star-forming galaxies in which our compact objects form, we sample  $M_*$, SFR and the average galaxy metallicity from three different distributions, based on observational scaling relations.  

\subsubsection{Stellar mass function of star-forming galaxies}
\label{sec:gsmf}

\begin{figure}
    \centering
    \includegraphics[width = 0.45\textwidth]{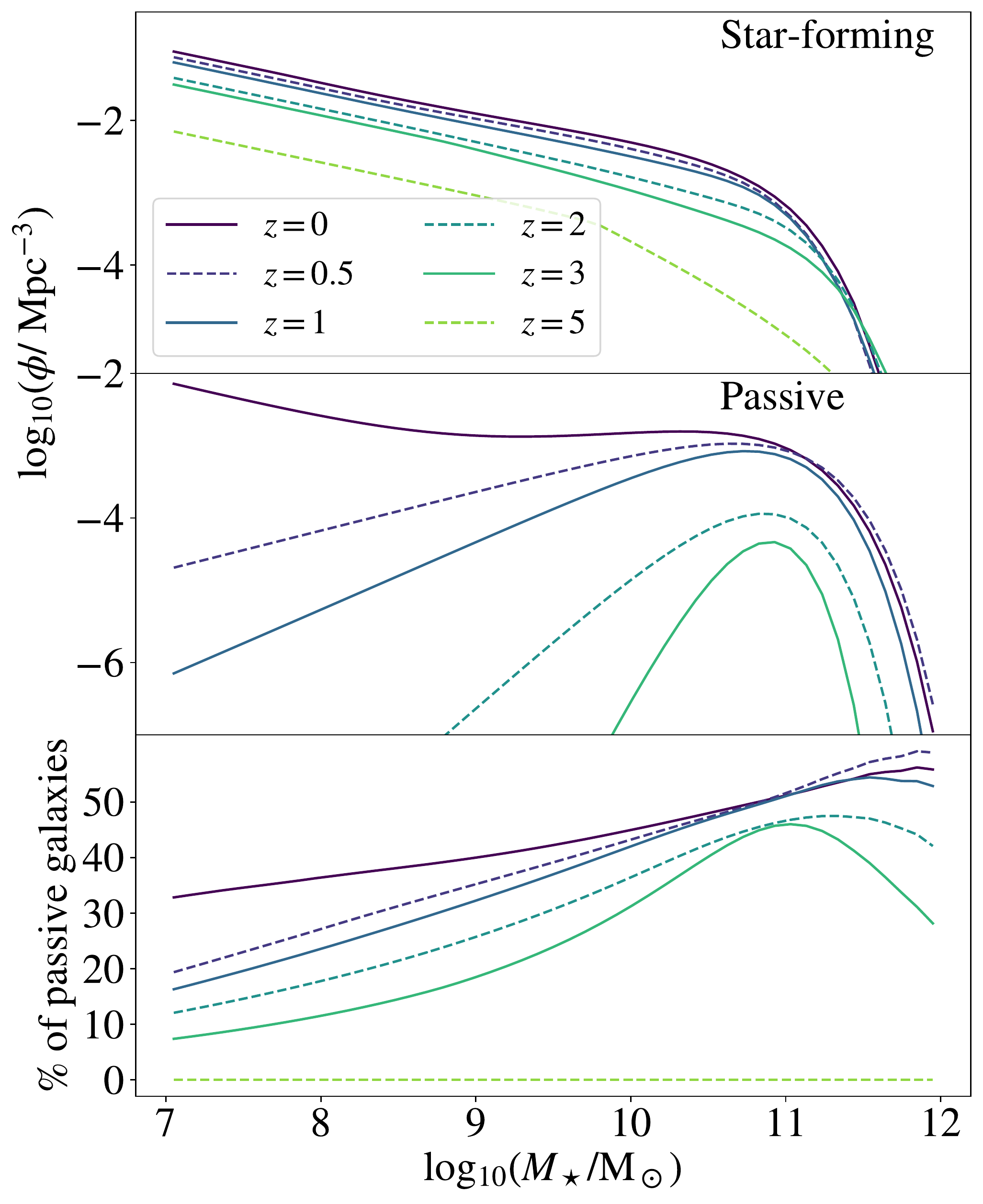}
    \caption{Upper panel: GSMF of star-forming galaxies from \protect\cite{chruslinska2019}, shown at different redshifts. Middle panel: GSMF of passive galaxies from \protect\cite{ilbert2013}. Lower panel: percentage of passive galaxies per stellar mass bin. 
    }
    \label{fig:gsmf}
\end{figure}

The stellar mass function of star-forming  galaxies is given in its simplest form as a Schechter  function \citep{schechter}:
\begin{equation}
\label{eq:gsmf}
    \phi(M_*, z)\,{}dM_*= \phi^*(z)\,{} e^{-M_*/M_{{\rm{cut}}}(z)} \left ( \frac{M_*}{M_{\rm{cut}}(z)}\right)^{\alpha_{\rm{GSMF}}}dM_*,
\end{equation}
where $\phi^*(z)$ is the normalisation, and  $M_{\rm{cut}}(z)$ 
is the stellar mass at which the Schechter function bends, changing from a single power law with slope $\alpha_{\rm{GSMF}}$ at low masses to an exponential cut-off at high masses. 
We adopted the galaxy stellar-mass function (GSMF) derived in \cite{chruslinska2019}, where the authors made an average of various GSMFs defined in the same redshift bin (see Table 1 of \citealt{chruslinska2019}). 
Figure~\ref{fig:gsmf} (upper panel) shows the resulting GSMF.

At each formation redshift, we sample a number of star-forming galaxies proportional to the galaxy number density, defined as the numerical integral of the GSMF between a minimum and a maximum stellar mass. In this way, similarly to what happens in cosmological simulations, we fixed the comoving volume $V= (100\,{}{\rm cMpc})^3$. We choose this volume as a good compromise between the computational cost of our models and the statistics of high-mass galaxies in the box. 
In Appendix \ref{sec:1gpc_vol}, we discuss the impact of this choice  
on our results. 

In the sampling procedure, we assume a maximum stellar mass 
$M_{\rm{max}} = 10^{12} $ M$_\odot$. 
The minimum stellar mass of the GSMF is highly uncertain, especially at high redshift \citep{conselice2016}. 
To this regard, we varied the minimum stellar mass $M_{\rm{min}} = 10^{6}, 10^{7}, 10^{8} $~M$_\odot$ to explore the impact of this parameter on our results. In Appendix \ref{sec:min_mass}, we show that the minimum mass $M_{\rm{min}}$ has a mild impact on the merger rate density in the local Universe. For the following results, we adopted $M_{\rm{min}} = 10^{7} $ M$_\odot$ as a fiducial value. We consider the redshift range between $z=0$ and 8. 
At higher redshift, the observations are too scanty to confidently extrapolate the GSMF.

\subsubsection{Star formation rate (SFR)}
\label{sec:sfr}

Galaxies can be classified in three main groups based on their SFR. (i) The majority of galaxies belong to the so-called \emph{galaxy main sequence}: 
they follow a tight relation between stellar mass and SFR at any redshift \citep{daddi2007, speagle2014, rodighiero2015, schreiber2015,   pantoni2019, lapi2020,leja2021, popesso2022}. 
(ii) A second group of galaxies, the \emph{starburst} galaxies, have higher SFR with respect to the main sequence \citep{rodighiero2011, caputi2017}. The main sequence and starburst galaxies form the population of star-forming galaxies. (iii) The third group of galaxies consists in the \emph{passive} (or quenched) galaxies, which have, on average, a lower SFR with respect to the main sequence \citep{renzini2015, bisigello2018, Santini2021}.   At low redshift, passive galaxies are the dominant population: they contain up to $\sim{70}$\% of the total stellar mass in the local Universe \citep{moffett2016}. 


In \galaxyrate{}, we describe the star-forming galaxy distribution in SFR
at fixed redshift and stellar mass 
with a double log-normal distribution 
\citep{daddi2007, rodighiero2011, sargent2012, bethermin2012, schreiber2015, ilbert2015, schreiber2015}. One of the two log-normal 
distributions is centered on the galaxy main sequence, the other 
on the starburst sequence \citep{boco2021}:  


\begin{equation}
\label{eq:2gauss}
\begin{aligned}
\mathcal{P}(\log_{10}{\rm{SFR}}| M_*, z) = A_{\rm{MS}}\exp {-\frac{(\log_{10} {\rm{SFR}} - \braket{\log_{10} {\rm{SFR}}} _{\rm{MS}})^2}{2\sigma_{\rm MS}^2}} + \\ A_{\rm{SB}}\exp{-\frac{(\log_{10} {\rm{SFR}} - \braket{\log_{10} {\rm{SFR}}}_{\rm{SB}})^2}{2\sigma_{\rm{SB}} ^2}},  
\end{aligned}
\end{equation}
where $\langle{}\log_{10}{\rm{SFR}}\rangle{}_{\rm MS}$ ($\langle{}\log_{10}{\rm{SFR}}\rangle{}_{\rm SB}$) is the average SFR of the main sequence (starburst galaxies), $\sigma_{\rm MS}$ ($\sigma_{\rm SB}$) is the standard deviation of the main sequence (starburst) galaxies, $A_{\rm{MS}} = 0.97$ and  $A_{\rm{SB}} = 0.03$ \citep{sargent2012}. We define the starburst sequence as $\braket{\log_{10} {\rm{SFR}}}_{\rm SB } = \braket{\log_{10} {\rm{SFR}}}_{\rm MS} + 0.59$ with $\sigma_{\rm{SB}} = 0.243$ dex \citep{sargent2012}. These values are obtained expressing the SFR in M$_\odot$ yr$^{-1}$. 

We compared two definitions of the galaxy main sequence $\braket {\log_{10}  {\rm{SFR}}  }_{\rm{MS}}$: the definition given in  \citet[][hereafter S14]{speagle2014} 
and \citet[][hereafter B18]{boogaard2018}. 
The definition of galaxy main sequence by S14 is
\begin{equation}
\label{eq:ms_speagle}
\begin{aligned}
\braket{\log_{10} {\rm{SFR}}(M_*,t)} = (0.84 - 0.026\,{}t) \log_{10} M_* - 6.51 - 0.11 \,{}t,
\end{aligned}
\end{equation}
where $t$ is age of the Universe in Gyr and $M_\ast$ the galaxy stellar mass in M$_\odot$. The definition of B18 is 
\begin{equation}
\label{eq:boogaard}
\braket{\log_{10} {\rm{SFR}}(M_*,z)} = 0.83\log_{10}\left(\frac{M_*}{M_0}\right) -0.83+1.74\log_{10}\left(\frac{1+z}{1+z_0}\right),
\end{equation}
where $M_0 = 10^{8.5}$ M$_\odot$ and $z_0 = 0.55$. 
Equation~\ref{eq:boogaard} has been estimated considering the Chabrier IMF  \citep{Chabier2003}; thus, we 
shifted the SFR normalisation to make it consistent with our default Kroupa IMF (\citealt{kroupa2001}; see Table B1 in \citealt{chruslinska2019} for more details about the conversion). 
The dispersion around the main sequence is another debated quantity. We 
assume $\sigma_{\rm{MS}} = 0.188$~dex for S14  
\citep{sargent2012} and $\sigma_{\rm{MS}} = 0.3$~dex for B18 \citep{chruslinska2019}.

\begin{figure}
    \centering
    \includegraphics[width = 0.45\textwidth]{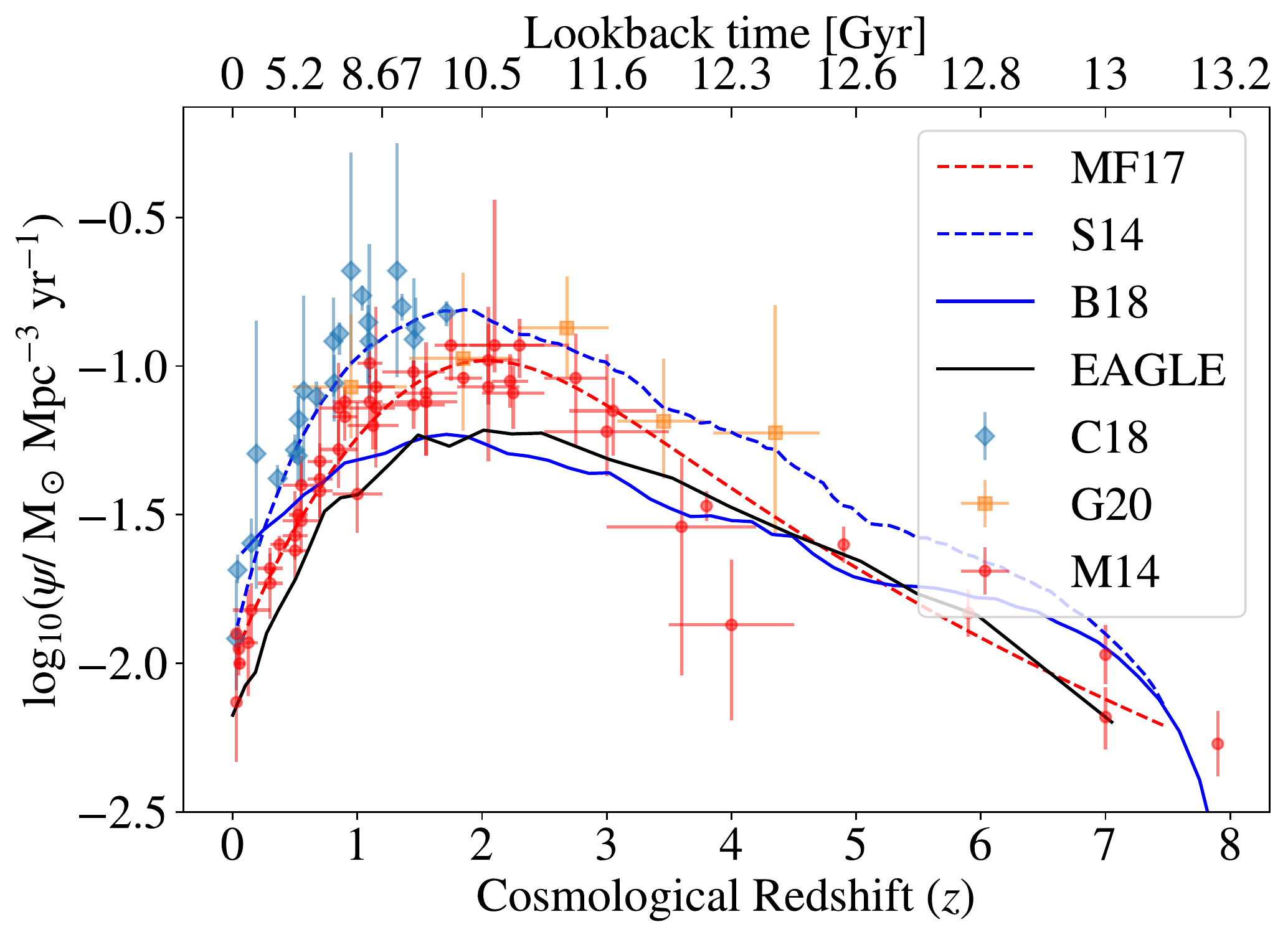}
    \caption{Cosmic SFR density, $\psi(z)$, as a function of redshift (lower $x-$axis) and look-back time (upper $x-$axis). The dashed and solid blue lines 
    indicate the star-forming main sequence from \protect\citet[][S14]{speagle2014}  and \protect\citet[][B18]{boogaard2018}, respectively. 
The black line is the SFR density from the {\sc eagle} box of (100~cMpc)$^3$ \protect\citep{schaye2015}. 
The red dots are from Figure 9 of \protect\citet[][M14]{madau2014}  and the red line is the fit to the cosmic SFR density by \protect\citet[][MF17]{madau2017}. 
The light blue dots are data points from \protect\citet[][C18]{casey2018}  and the orange dots from \protect\citet[][G20]{gruppioni2020}. 
All the data have been re-normalized to assume a Kroupa IMF \protect\citep{kroupa2001}.}
    \label{fig:sfrd}
\end{figure}

After sampling a number of galaxies from  the GSMF and 
assigning them the SFR as described above, we can derive the total cosmic SFR density, by simply evaluating the average SFR over the population of galaxies and multiplying it by the galaxy number density at each  redshift. Figure \ref{fig:sfrd} shows different resulting cosmic SFR densities depending on the definition of the main sequence, and compares them to observational data. 
The star formation history we obtain adopting the results of S14 is in agreement with some of the most recent data points \citep{casey2018} and is slightly higher than the fit by \cite{madau2017}. In contrast, the results of B18 lie below most observational data in the redshift range $z\in[1,3]$, across cosmic noon. Hence, the models by S14 and B18 allow us to 
compare two different evolutions of the star formation history, in terms of normalisation and redshift evolution.    
In the same figure, we also show the cosmic SFR density 
from the {\sc eagle} cosmological simulation  with a (100~cMpc)$^3$ volume \citep{furlong2015}. 
The cosmic SFR density obtained adopting the results of B18  
is close to the one evaluated with the {\sc eagle} 
especially at high redshift ($z>1$).

\subsubsection{Metallicity distribution}
\label{sec:metallicity}

\begin{figure}
    \centering
    \includegraphics[width = 0.45
    \textwidth]{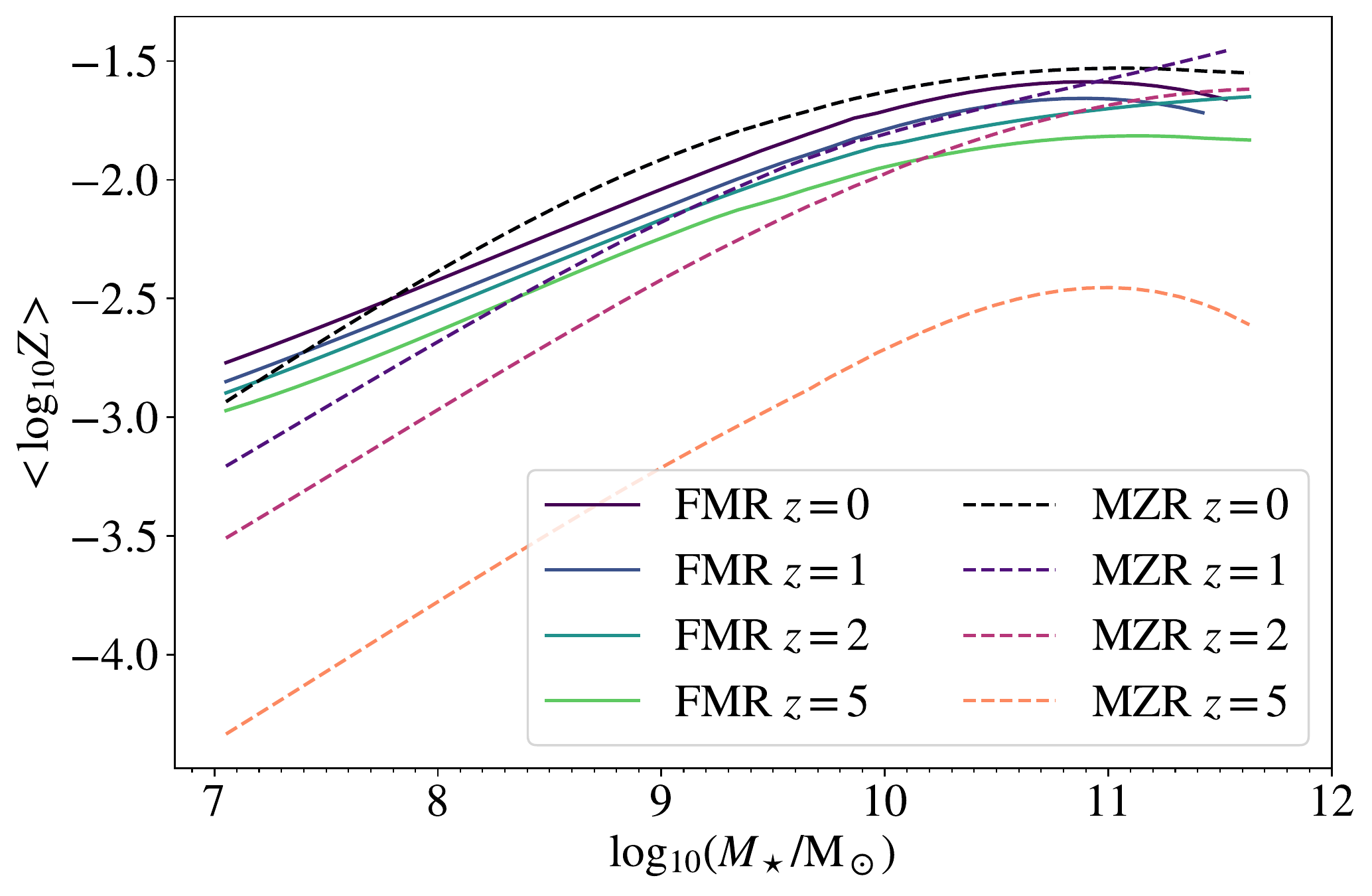}
    \caption{Average metallicity of star-forming galaxies obtained by sampling from the B18 main sequence and from the FMR  \protect{\citep[solid lines,][]{mannucci2011}}, compared with the MZR  \protect{\citep[dashed lines,][]{ chruslinska2019}}, at different redshifts.}
    \label{fig:met_comp}
\end{figure}

 For each galaxy, we  sample an average metallicity $Z$ from observational scaling relations.  
 Two main scaling relations have been proposed between the metallicity and other 
 galaxy properties: the mass metallicity relation \citep[MZR, ][]{tremonti2004,Kewley2008,maiolino2008, mannucci2009, magnelli2012, zahid2014, genzel2015, sanders2020} and the fundamental metallicity relation \citep[FMR,][]{mannucci2010, mannucci2011, hunt2012, hunt2016, curti2020, sanders2020}. 
The MZR is a correlation between metallicity and stellar mass, based on observations of galaxies with stellar mass in the range  $10^8-10^{12}$~M$_\odot$ and redshift $z \sim 0-3.5$ \citep{tremonti2004,zahid2014}. 
In general, at fixed stellar mass the MZR predicts a steep decline in $Z$ towards higher redshifts. Some earlier works \citep[e.g.,][]{maiolino2008, mannucci2009, magnelli2012} found a slow evolution of the MZR out to $z\sim 2$ but a very sharp decline 
of about $0.4-0.5$ dex in $Z$ 
between $z=2.5$ and $3.5$. 
Extrapolating this trend at $z>3.5$, we expect a rapid decrease of the average galaxy metallicity in the early Universe.

The FMR is a three-parameter relation among stellar mass, SFR and metallicity. 
 According to the FMR, the metallicity decreases at increasing SFR, for a fixed stellar mass \citep{mannucci2010, hunt2016}. 
The FMR is thought to be almost redshift independent\footnote{Recent data from the Early Release Observations program of the James Webb Space Telescope  suggest that the FMR might hold up to redshift $z\sim{}8$ \protect\citep{Curti2022}.}, as confirmed by observations out to $z\sim 3.5$ \citep{mannucci2010}. The redshift $z$ is not a parameter of the FMR, and the metallicity evolution with redshift at fixed stellar mass is entirely explained with the redshift evolution of the SFR described by the main sequence (see Section \ref{sec:sfr}). 
This results in a shallow decline of $Z$ with redshift at fixed stellar mass. Therefore, while the redshift evolution of the FMR and MZR at $z \lesssim 2$ is similar, 
the evolution of the two relations becomes completely different at $z \gtrsim 3$ (Fig.~\ref{fig:met_comp}). 
We refer to Section~3 of \cite{boco2021} for a thorough discussion about the tension between the two metallicity distributions.

In the following, we will consider both MZR and FMR.  To implement the MZR, we use the same definition as in Section 3.2 of \cite{chruslinska2019}. 
For the FMR, we use the definition in Equation~2 of  \cite{mannucci2011}.

Both metallicity relations are usually expressed in terms of the relative abundance of oxygen and  
hydrogen, $12 + \log_{10}(\rm{O/H})$.  We need to convert this quantity to the total mass fraction of heavy elements $Z$, since our stellar-evolution models depend on $Z$ (see Section \ref{sec:pop-synth}). 
We assumed $Z_\odot$ = 0.0153 and $12 + \log_{10} ({\rm{O/H}})_\odot = 8.76$ \citep{caffau2011}. Other choices of the solar metallicity value result in a different normalization of the merger rate density curve (Appendix~\ref{sec:met_sol}).

Observations show that at a given stellar mass, there is an intrinsic scatter around both the MZR and FMR \citep{tremonti2004, Kewley2008, zahid2014,chen2022}. Thus, we assumed a log-normally distributed scatter around the mean metallicity given by both the MZR and FMR. We considered a spread $\sigma_0 = 0.15$ dex for both relations. 
Furthermore, we calculate the metallicity of single stars assuming a metallicity gradient inside each galaxy, that is the metallicity of single stars 
is log-normally  distributed around the mean metallicity of the galaxy with $\sigma_{1} = 0.14$ dex (\citealt{chruslinska2019}; see also \citealt{sanchez2014} and \citealt{sanchez2016}). 

Figure \ref{fig:met_comp} shows the MZR and the FMR averaged over the SFR for six different redshift bins. While at $z\lesssim{2}$ the two relations give almost identical results, at $z\gtrsim{3}$ the MZR yields a rapid evolution of the metallicity, resulting in much lower values than those obtained from the FMR. 


\begin{figure*}
    \centering
    \includegraphics[width = 0.90\textwidth]{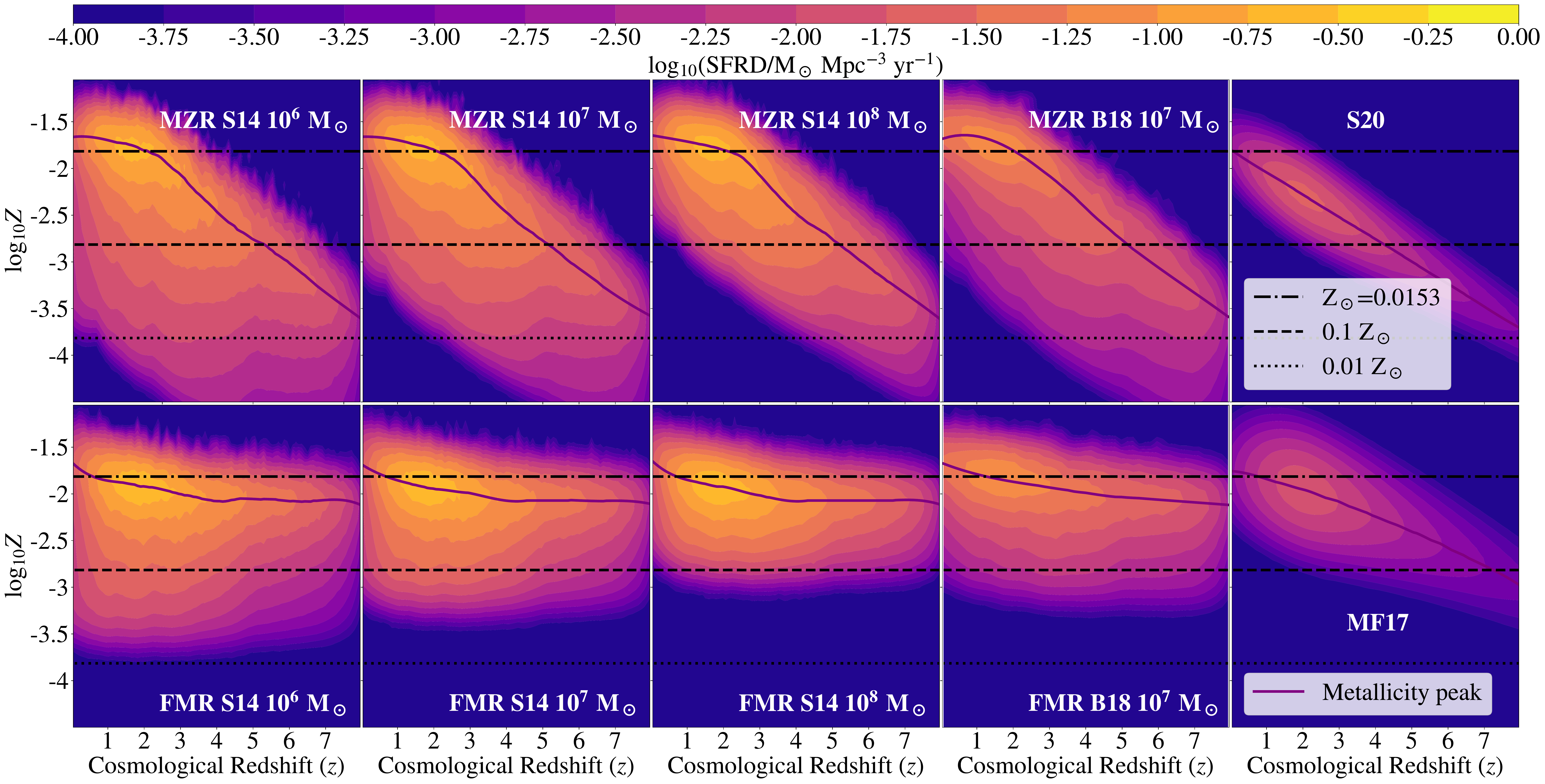}
    \caption{Distribution of the SFR density per metallicity bin, SFRD$(z,Z)$,  resulting from \galaxyrate{} (Section \ref{sec:galaxymodel}) and \cosmorate{} (Appendix \ref{sec:cosmorate}). Starting from the left: the first, second and third columns show the SFRD$(z,Z)$ obtained with S14, where we varied the minimum sampled stellar mass from the GSMF: $M_{\rm min} = 10^6, 10^7$ and $10^8$ M$_\odot$, respectively. The fourth column shows the SFRD$(z,Z)$ obtained with B18 and $M_{\rm min} = 10^7$ M$_\odot$. The four upper (lower) leftmost panels show the results obtained with the MZR (FMR). The rightmost column shows the SFRD$(z,Z)$ obtained with \cosmorate{}, adopting the average metallicity evolution described in S20 and MF17 in the upper and lower panel, respectively. The purple solid line in all panels shows the metallicity associated with the highest value of the SFRD$(z,Z)$ as a function of redshift.}
    \label{fig:sfrd_zZ}
\end{figure*}

Figure~\ref{fig:sfrd_zZ} shows the resulting metallicity-dependent SFR density, SFRD$(z,Z)$, for each of the aforementioned models. 
Both  the FMR and MZR produce a non-negligible fraction of metal-poor   star formation ($\log_{10} Z\leq -3$), both at high and low redshift. The fraction of metal-poor star formation increases if we assume lower values of $M_{\rm min}$, because smaller galaxies are also more metal poor. 
The metallicity associated with the highest value of the SFRD$(z,Z)$ drops at high redshift if we assume the MZR or the metallicity model in \cite{santoliquido2020}, while it decreases very mildly for the FMR, in agreement with \cite{chruslinska2021} and \cite{boco2021}.

\subsubsection{Passive galaxies}
\label{sec:passive}

In our model, we generate the mass distribution of passive galaxies 
from the GSMF by \cite{ilbert2013}, 
by adopting a double Schechter function \citep{pozzetti2010}:
\begin{equation}
 \phi(M_*)\,{}dM_* = e^{-\frac{M_*}{M_{\rm{cut}}}}\left[ 
 \phi_1 \left ( \frac{M_*}{M_{\rm{cut}}}\right)^{\alpha_1} + \phi_2 \left( \frac{M_*}{M_{\rm{cut}}} \right)^{\alpha_2}\right]  \frac{dM_*}{M_{\rm{cut}}},  
\end{equation}
where the parameters $M_{\rm{cut}}, \phi_1, \alpha_1, \phi_2$ and $\alpha_2$ depend on redshift and are defined as in Table 2 of \cite{ilbert2013}. We linearly interpolated them  between redshift bins at $z\leq3$. 
\cite{ilbert2013} show that the number density of passive galaxies rapidly drops at $z\sim{3}$. Hence, in our model we assume that the number of passive galaxies is zero at $z^{\rm{pass}}_{\rm{max}}=3$. 

The middle panel of Figure~\ref{fig:gsmf} shows the stellar mass density of passive galaxies for some redshift bins. The lower panel of the same Figure shows the percentage of passive galaxies with respect to star-forming galaxies as a function of mass for the same redshift bins. Passive galaxies represent the majority of galaxies in the local Universe at high mass ($M_* > 10^{11}$~M$_\odot$, \citealt{moffett2016, mcleod2021}). 
There are several ways in the literature to define passive galaxies and the resulting population can show different properties based on the chosen definition \citep[see, e.g., ][]{donnari2021}. 
In our fiducial definition, we assume that passive galaxies have a value of the SFR at least one dex below the main sequence, for a given stellar mass \citep{Donnari2019}. 
For each passive galaxy, we extract a SFR value uniformly distributed between SFR$_{\rm min}^{\rm pass}$ and SFR$_{\rm max}^{\rm pass}$. We assume a fixed value for SFR$_{\rm min}^{\rm pass} = 10^{-4}$~M$_\odot$~yr$^{-1}$, since it is unlikely that passive galaxies with $M_*> 10^8$~M$_\odot$ have lower SFR \citep{renzini2015}. 
We assume $\log_{10}{\rm SFR}_{\rm max}^{\rm pass} = \langle{}\log_{10}{\rm{SFR}}\rangle{}_{\rm MS} - N_{\rm{pass}}\,{} \sigma_{\rm{MS}}$, where $\langle{}\log_{10}{\rm{SFR}}\rangle{}_{\rm MS}$ is given by either Equation \ref{eq:ms_speagle} (S14) or Equation \ref{eq:boogaard} (B18) depending on which main sequence we assume,  $\sigma_{\rm{MS}}$ is the dispersion around the main sequence, and $N_{\rm{pass}} = {\rm{int}}(1/\sigma_{\rm{MS}})$. Thus in the case of S14 (B18) $N_{\rm{pass}} = 5$ (3). With this definition of ${\rm SFR}_{\rm max}^{\rm pass}$, passive galaxies have maximum SFR  $\approx{1}$ dex below the main sequence. We estimate the average metallicity of passive galaxies either with the MZR or the FMR.

Here, we assume for simplicity that a passive galaxy cannot be the FG of the stellar progenitors of BCO mergers. 
This is a reasonable assumption, because the total SFR happening in passive galaxies in the local Universe is $< 3$ \% of the total SFR. 
However, BCO mergers can take place in passive galaxies, because the delay time $t_{\rm del}$ between the formation and the merger of a binary system can be several Gyr long: the FG might become a passive galaxy during $t_{\rm del}$ and/or might  merge together with a passive galaxy.


\subsection{Merger rate density}\label{sec:MRD}
To evaluate the merger rate density we proceed in a similar way as in \cite{mapelli2017} (see also \citealt{mapelli2018,artale2019}).   
At each time step $\Delta{}t$, we associate to each galaxy in our sample a total mass $M_{\rm new}(z)={\rm SFR}(z)\,{}\Delta{}t$ of newly formed stars, 
by randomly sampling a population of stars with total zero-age main sequence mass 
$M_{\rm new}$ from our population synthesis simulations. The metallicity distribution of these randomly sampled stars is the same as the metallicity distribution of the entire FG\footnote{This might overestimate the contribution of metal-poor stars
, as newly born stars are generally more metal-rich than previous star formation episodes \citep[e.g.,][]{Peeples2014}.} at redshift $z$. 
  This procedure allows us to associate the BCOs that form from this stellar population in our {\sc mobse} simulations to their FG at a given formation redshift. Hereafter, we indicate with $t_{\rm form}$ ($z_{\rm form}$) the time (redshift) at which the progenitor stars of a BCO are associated with their FG.
Each BCO merges at a time $t_{\rm merg}$, which is estimated as $t_{\rm merg}=t_{\rm form}-t_{\rm del}$. Here, both $t_{\rm merg}$ and $t_{\rm form}$ are expressed in terms of look-back times, while $t_{\rm del}$ is the delay time. 

To evaluate the global cosmic  merger rate density at a given redshift $z_{\rm merg}$, 
we simply divide the total number of mergers happening at $z_{\rm{merg}}$ in all the considered galaxies, by the total comoving volume $V$. 
For each considered galaxy, we also calculate 
a merger rate per galaxy $n_{\rm GW}$, i.e. the number of compact objects merging in that galaxy per unit time. 


\subsection{Host galaxy (HG)}
\label{sec:mergergalaxies}

\begin{figure*}
    \centering
    \includegraphics[width = 0.90\textwidth]{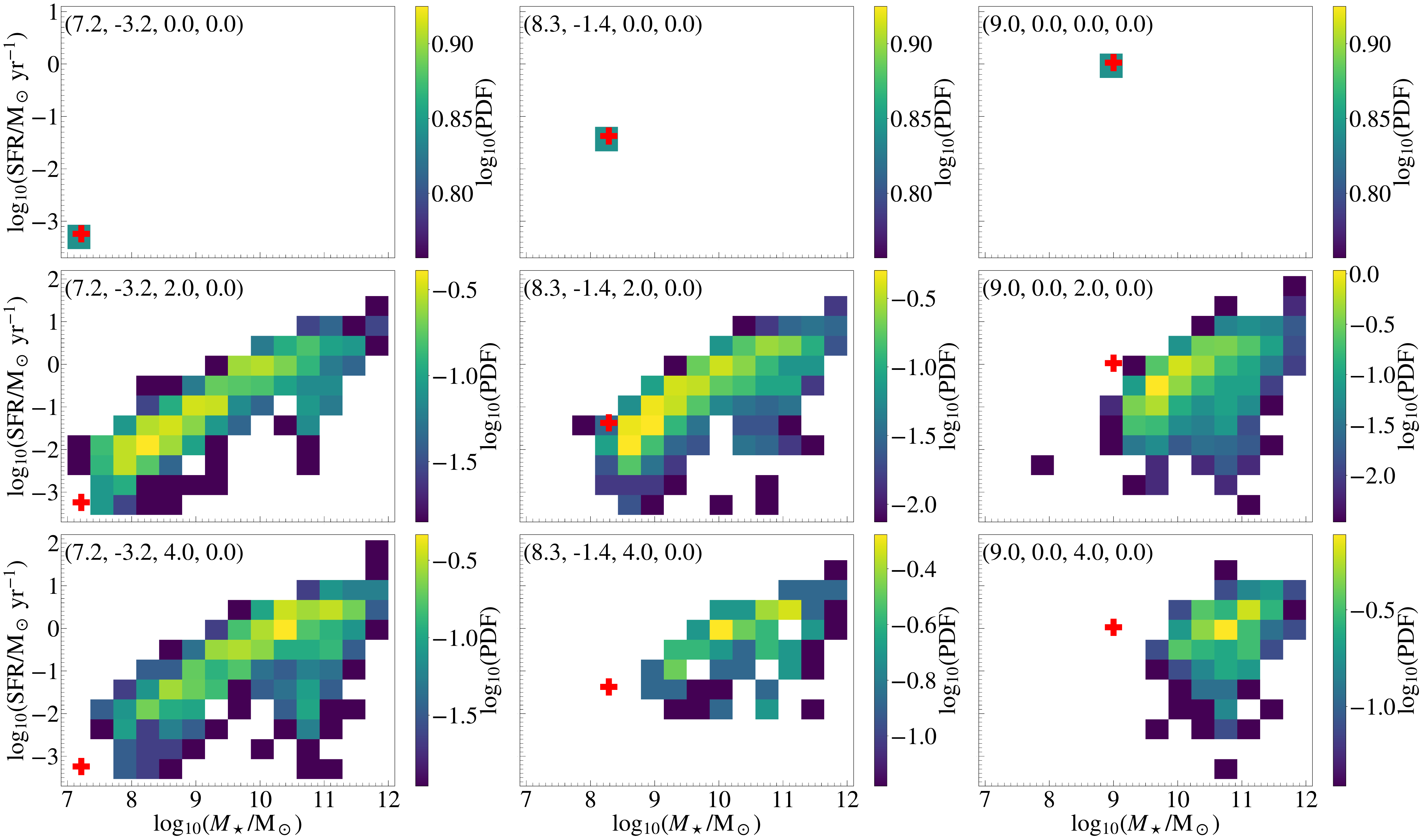}
    \caption{Conditional probability  $\mathcal{P}({\rm{M}}_{\rm{host}}, {\rm{SFR}}_{\rm{host}}| {\rm{M}}_{\rm{form}}, {\rm{SFR}}_{\rm{form}},z_{\rm{form}}, z_{\rm{merg}})$ evaluated adopting the merger tree of the {\sc{eagle}} cosmological simulation, for various properties of the FGs, annotated at the top of each panel following the order  $(\log_{10}({\rm{M_{form}}}/{\rm{M}}_\odot), \log_{10}({\rm{SFR_{form}}}/{\rm{M}}_\odot{\rm{yr}}^{-1})
    , z_{\rm{{form}}}, z_{\rm{merg}})$. The  red cross marks the location of the FG at $z_{\rm{form}}$. 
    In this plot $z_{\rm{merg}} = 0$.}
    \label{fig:prob}
\end{figure*}


According to our definition, the HG of a BCO is the galaxy where the BCO merges. The HG can be the same as the FG, if the binary system forms and merges in the same galaxy, but it can also be different from the FG, if for example the FG has undergone galaxy-galaxy mergers by the time the two compact objects reach coalescence. Furthermore, even if the BCO forms and merges in the same galaxy, the mass, SFR and metallicity of the galaxy might change significantly after $t_{\rm del}$. 

To reconstruct the merger history of galaxies
in \galaxyrate{} we used the information from a  merger tree. 
A merger tree encodes the entire assembly history and property evolution of each single galaxy across cosmic time \citep[e.g.,][]{McAlpine2016,Qu2017}. 
Here, we show the results of the merger trees taken  from the {\sc eagle} cosmological simulation \citep{schaye2015}, but \galaxyrate{} can use any other possible merger trees obtained from cosmological simulations or semi-analytical models. 


One of the main purposes of our new approach is to build a fast tool, thus we compressed the information contained in multiple merger trees, by evaluating the conditional probability 
$\mathcal{P}(M_{\rm host}, {\rm{SFR}}_{\rm host}| M_{ \rm form}, {\rm{SFR}}_{\rm form}, z_{\rm form}, z_{\rm merg})$, i.e. the probability that the HG of the BCO merger has mass $M_{\rm host}$ and star formation rate ${\rm{SFR}}_{\rm host}$, given the mass ($M_{ \rm form}$) and SFR (${\rm{SFR}}_{\rm form}$) of the FG, and given the formation ($z_{\rm form}$) and merger redshift ($z_{\rm merg}$) of the BCO. 
We evaluate  the conditional probability 
$\mathcal{P}(M_{\rm host}, {\rm{SFR}}_{\rm host}| M_{ \rm form}, {\rm{SFR}}_{\rm form}, z_{\rm form}, z_{\rm merg})$ in an empirical way, directly from the merger trees. %
We count the number of galaxies at a fixed $z_{\rm merg}$ that are inside each bin in the $(M_{\rm host}, {\rm{SFR}}_{\rm host})$ plane, by keeping fixed the condition on $M_{\rm form}$, ${\rm{SFR}}_{\rm form}$ and $z_{\rm form}$. Figure~\ref{fig:prob} shows some examples of this conditional probability, in which we compare different values of the formation redshift $z_{\rm{form}}$ 
and different properties of the FGs. 
If the FG has no time to evolve between the formation and merger of the BCO (short $t_{\rm del}$), the properties of the HG remain the same (upper row) as those of the FG, while if the FG has more time to evolve (long $t_{\rm del}$) then the HG can be very different from the FG (lower rows).

Once we have defined the conditional probability, we sample one HG for each FG. 
Thereafter, we link each sampled HG from the merger trees to one and only one galaxy obtained through the observational scaling relations, 
considering both star-forming galaxies (Section \ref{sec:galaxymodel}) and passive galaxies (Section \ref{sec:passive}).   
Finally, to calculate the merger rate per galaxy at a given merger redshift $z_{\rm merg}$, we sum up all the mergers of compact objects formed at any $z_{\rm form}$ and merging at $z_{\rm merg}$ in the same galaxy.

The conditional probabilities in Figure~\ref{fig:prob} show that, in some cases, the stellar mass of the HG might be smaller than the stellar mass of the FG. This happens for two reasons. In most cases, different physical mechanisms such as strong galactic outflows or galaxy interactions producing tidal stripping can reduce the stellar mass of the galaxies \citep[see e.g.,][]{MacLow1999,Efstathiou2000,Hopkins2012}. In other 
cases, 
the subfind algorithm 
might misclassify the star particles belonging to a given galaxy. This issue is more common at the pericenter of a subhalo orbiting a larger halo  \citep[see e.g.,][]{Muldrew2011,Knebe2011}.

\begin{table*}
    \centering
    \begin{tabular}{l|l|l|l}
    \toprule
        Section & Parameter/Model & Value(s)/Choice(s) & Reference(s) \\
         \hline
         \ref{sec:pop-synth} & Core Collapse SN & Rapid & \cite{fryer2012} \\ 
                            & Natal Kicks & $v_{\rm{kick}} \propto m_{\rm{ej}}/m_{\rm{rem}}v_{\rm{H05}}$ 
                            & \cite{giacobbo2020}\\
                            & $\alpha$ Common Envelope & 1, 3 and 5 & \cite{Webbink1984}\\
                            & $\lambda$ Common Envelope & Depends on star properties & \cite{claeys2014} \\
                            & Primary star IMF & Kroupa with $M \in [5, 150]$ M$_\odot$ & \cite{kroupa2001}\\
                            & Mass ratio & $\mathcal{F}(q) \propto q^{-0.1} $ & \cite{sana2012}  \\
                            & Orbital period & $\mathcal{F}(\Pi) \propto \Pi^{-0.55} $ & \cite{sana2012} \\
                            & Eccentricity & $\mathcal{F}(e) \propto e^{-0.42} $ & \cite{sana2012}\\
                            & Progenitor metallicity & $Z \in [0.0002, 0.02]$ & \cite{giacobbo2018b} \\
         \hline 
        \ref{sec:gsmf} & Star-forming GSMF & Single Schechter & \cite{chruslinska2019}\\
                        & $M_{\rm{min}}$ & $10^{6}, 10^{7}, 10^{8}$ M$_\odot$ & \cite{conselice2016} \\
                        & $\alpha_{\rm{GSMF}}$ & constant, varying with $z$ & \cite{chruslinska2019} \\
                         & $z_{\rm{max}}$ & 8 & \cite{ilbert2013}\\
         \hline
         \ref{sec:sfr} & $A_{\rm{MS}}$ & $0.969^{+0.004}_{-0.006}$ & \cite{sargent2012} \\
                      & $\langle{}\log_{10}{\rm{SFR}}\rangle{}_{\rm{MS}}$ & S14; B18 & \cite{speagle2014, boogaard2018} \\
                      & $\sigma_{\rm{MS}}$ & $0.188^{+0.003}_{-0.003}$, 0.3 & \cite{sargent2012, chruslinska2019}\\
                      & $A_{\rm{SB}}$ & $0.031^{+0.006}_{-0.004}$ & \cite{sargent2012}\\
                     &$\langle{}\log_{10}{\rm{SFR}}\rangle{}_{\rm{SB}}$ & $\langle{}\log_{10}{\rm{SFR}}\rangle{}_{\rm{MS}}$ +  $0.59^{+0.06}_{-0.13}$ & \cite{sargent2012}\\
           &$\sigma_{\rm{SB}}$ & $0.243^{+0.078}_{-0.047}$ & \cite{sargent2012} \\
        \hline
        \ref{sec:metallicity} & Metallicity relation & MZR; FMR & \cite{chruslinska2019, mannucci2011}\\
                            & $Z_\odot$ & 0.0153 &\cite{caffau2011}\\
                            & $\sigma_0$ & 0.05; 0.15; 0.30  & \cite{boco2021} \\
                            & $\sigma_1$ & 0.00001; 0.14; 0.30  & \cite{chruslinska2019}\\
                            & Metallicity calibration & Photoionization
models; $T_e-$based & \cite{maiolino2008, curti2020} \\
        \hline
        \ref{sec:passive} & Passive GSMF & Double Schechter & \cite{ilbert2013}\\
                         & $z_{\rm{max}}^{\rm{pass}}$ & 3 & \cite{ilbert2013} \\
                        & ${\rm{SFR}}_{\rm{max}}^{\rm{pass}}$ & 1 dex below the adopted MS & \cite{Donnari2019} \\
                        & ${\rm{SFR}}_{\rm{min}}^{\rm{pass}}$ & $10^{-4}$ M$_\odot$ yr$^{-1}$ & \cite{renzini2015}\\
        \hline 
        \ref{sec:mergergalaxies} & Merger trees & {\sc{eagle}} & \cite{schaye2015} \\
        
    \bottomrule
    \end{tabular}
    \caption{Parameters and models adopted in this version of {\sc{galaxy}$\mathcal{R}$\sc{ate}}. S14 and B18 refer to \protect\cite{speagle2014} and \protect\cite{boogaard2018}, respectively. For the MZR we adopt the definition in \protect\cite{mannucci2009} and \protect\cite{chruslinska2019}. For the FMR, we adopt the definition in \protect\cite{mannucci2011}.}
    \label{tab:param}
\end{table*}

\section{Results}
\label{sec:results}

\subsection{Merger rate density}

\begin{figure*}
    \centering
    \includegraphics[width = 0.90\textwidth]{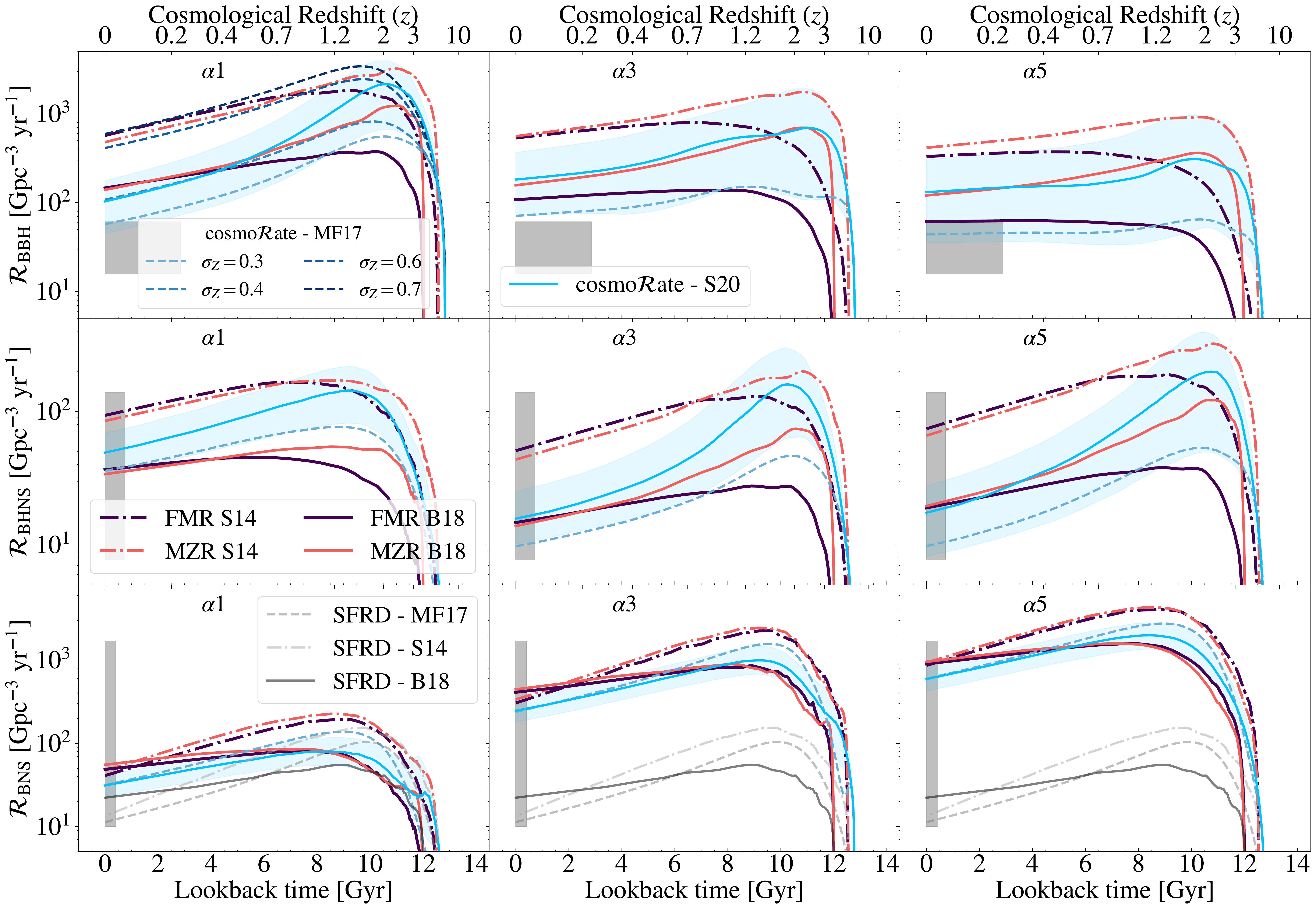}
    \caption{Evolution of the merger rate density $\mathcal{R}(z)$  in the comoving frame as a function of the look-back time (lower $x-$axis) and of the cosmological redshift (upper $x-$axis), for $\alpha=1,3,5$ (columns) and for BBHs, BHNSs and BNSs (rows).  
    Here, we show the results obtained 
    with $M_{\rm min}=10^7~\rm{M}_{\odot}$ for two different metallicity evolution models (MZR in red and FMR in blue)  and for two definitions of the star-forming main sequence (B18, solid lines, and S14, dash-dotted lines). The solid and  dashed light blue lines are the merger rate densities obtained with \cosmorate{}, assuming the average metallicity evolution in S20 and 
    MF17, respectively. For the model MF17 with $\alpha{}=1$, we evaluated the merger rate density of BBHs (upper left-hand panel) with different values of $\sigma_{Z} = 0.3, 0.4, 0.6 $ and $0.7$ (for more details, see Appendix \ref{sec:cosmorate}). In the other panels, we fix   $\sigma_{Z} = 0.3$ for the model M17. 
    The shaded light-blue areas are the 50\% credible intervals of the S20 model considering the uncertainties on  metallicity.  For BBHs, BHNSs and BNSs the gray shaded areas shows the 90\% credible interval of the local merger rate density \protect{\citep{pop3_2021}}. 
    The width of the shaded areas on the $x-$axis corresponds to the instrumental horizon obtained by assuming BBHs, BHNS, BNSs of mass $(30,\,{} 30)$, $(10,\,{}1.4)$ and $(1.4,\,{}1.4)$ M$_\odot$, respectively, and O3 sensitivity \protect{\citep{abbott201prospects}}. The dashed gray line in the lower panels is the total cosmic SFR density from MF17.  
    The solid and dashed-dotted gray lines are the cosmic SFR densities derived from B18 and S14 star-forming main sequences, respectively.}
    \label{fig:mrd}
\end{figure*}

Figure \ref{fig:mrd} shows the evolution of the merger rate density $\mathcal{R}(z)$ for three different values of the $\alpha$ parameter, for the two main sequence models (B18 and S14) and for the two metallicity relations (FMR and MZR). 
In this Figure, we also compare the results obtained with  {\sc{cosmo$\mathcal{R}$ate}}  (Appendix~\ref{sec:cosmorate}) and those obtained with 
the new code \galaxyrate{}.  
For each value of the common envelope efficiency $\alpha$, \galaxyrate{} 
produces a higher local merger rate density of BBHs and BHNSs with respect to the model MF17, that we obtained with {\sc{cosmo$\mathcal{R}$ate}} assuming the average metallicity evolution from \cite{madau2017} and a metallicity spread $\sigma{}_Z=0.3$. This 
springs from the interplay of two factors. First, BBHs and BHNSs merge more efficiently if they have metal-poor progenitors in our population-synthesis models (Appendix \ref{sec:delay}). Second, the SFRD$(z,Z)$ distribution  of 
\galaxyrate{} yields 
a longer tail of low-metallicity star formation at different redshifts with respect to \cosmorate{} (Figure \ref{fig:sfrd_zZ}).  The differences between the BBH
merger rate density obtained with \galaxyrate{} and \cosmorate{} (model M17) can
be reconciled by assuming a larger metallicity spread $\sigma{}_Z > 0.4$ in
\cosmorate{}.

The dependence of the BBH/BHNS merger rate on metallicity 
also appears when we vary both the main sequence definition and the metallicity relation. We have already seen that with B18 
the SFR density at redshift $z\sim 2$ is $\sim 0.5$ dex lower with respect to S14  (Figure \ref{fig:sfrd}). 
Thus, B18 
quenches the formation of BBHs and BHNSs at high redshift. 
As a result, the merger rate density of BBHs evaluated with B18  and $\alpha = 1$ is $\sim 4$ times lower with respect to S14 
in the local Universe. 


The merger rate density of BBHs and BHNSs in the local Universe is almost the same if we adopt the MZR or the FMR. However, the two scaling relations (MZR and FMR) produce a completely different slope of the merger rate density of BBHs at $ z \gtrsim 1$. 
 For $\alpha = 5$, the merger rate density of BBHs almost decreases 
 between $0<z<2$, 
 if we assume the FMR. This 
happens because the case with $\alpha = 5$ has the longest delay times (Appendix~\ref{sec:delay}) and the FMR yields only a mild decrease of the average metallicity 
with redshift (Figure \ref{fig:met_comp}).

In contrast, the merger rate density of BNSs does not depend on the adopted metallicity relation. 
Figure \ref{fig:mrd} (lower panel) 
shows the three cosmic SFR densities 
we took in account in this work. 
The merger rate density of BNSs 
approximately scales with the adopted cosmic SFR density, 
because in our population-synthesis models the BNSs are almost independent of metallicity and  their delay times are predominantly short 
\citep[e.g.,][]{dominik2012,dominik2013,klencki2018,mapelli2018,mapelli2019,neijssel2019,santoliquido2021}. 
In Appendix \ref{sec:min_mass}, we also discuss the impact of the minimum galaxy stellar mass on the merger rate density. 

The merger rate density of BNSs and BHNSs is within the 90\% credible interval estimated by the LVK with GWTC-3 \citep{pop3_2021} for all considered assumptions, while the merger rate density of BBHs predicted by our models is higher than the LVK range. This discrepancy is 
 the consequence of several factors. Firstly, here we assumed that all BBHs form via unperturbed binary evolution. Considering alternative channels might significantly affect the merger rate evolution \citep[e.g.,][]{mapelli2022}. Furthermore, current binary evolution models might overestimate the merger rate of BBHs and its dependence on metallicity, because of the large uncertainties they are affected by \citep[e.g.,][]{gallos_garcia2021, belczynski2022}. Finally, the measurement of gas and star metallicity is  affected by large uncertainties, especially at high redshift  \citep[e.g.,][]{maiolino2019}. 

The merger rate density of BCOs likely extends to a redshift higher than the maximum value we considered in this work ($z_{\rm{max}}= 8$). Here, we do not consider $z > 8$ because we prefer to avoid to arbitrarily extrapolate the observational relationships at higher redshift. Furthermore, the merger rate density at $z\gtrsim{}8$ is likely dominated by population III stars \citep[e.g.,][]{inayoshi2017,liu2021, ng2021, tanikawa2021}, which are not included in our formalism. We will include them in a follow-up study.


\subsection{Formation and host galaxies across cosmic time}

In this Section, we look at the distribution of HG properties, namely stellar mass, SFR and metallicity. We compare them with the distribution of FGs. 
For the sake of brevity, we show the results only for the definition of galaxy main sequence in B18. 
\subsubsection{Stellar Mass}

\begin{figure*}
    \centering
    \includegraphics[width = 
    0.85\textwidth]{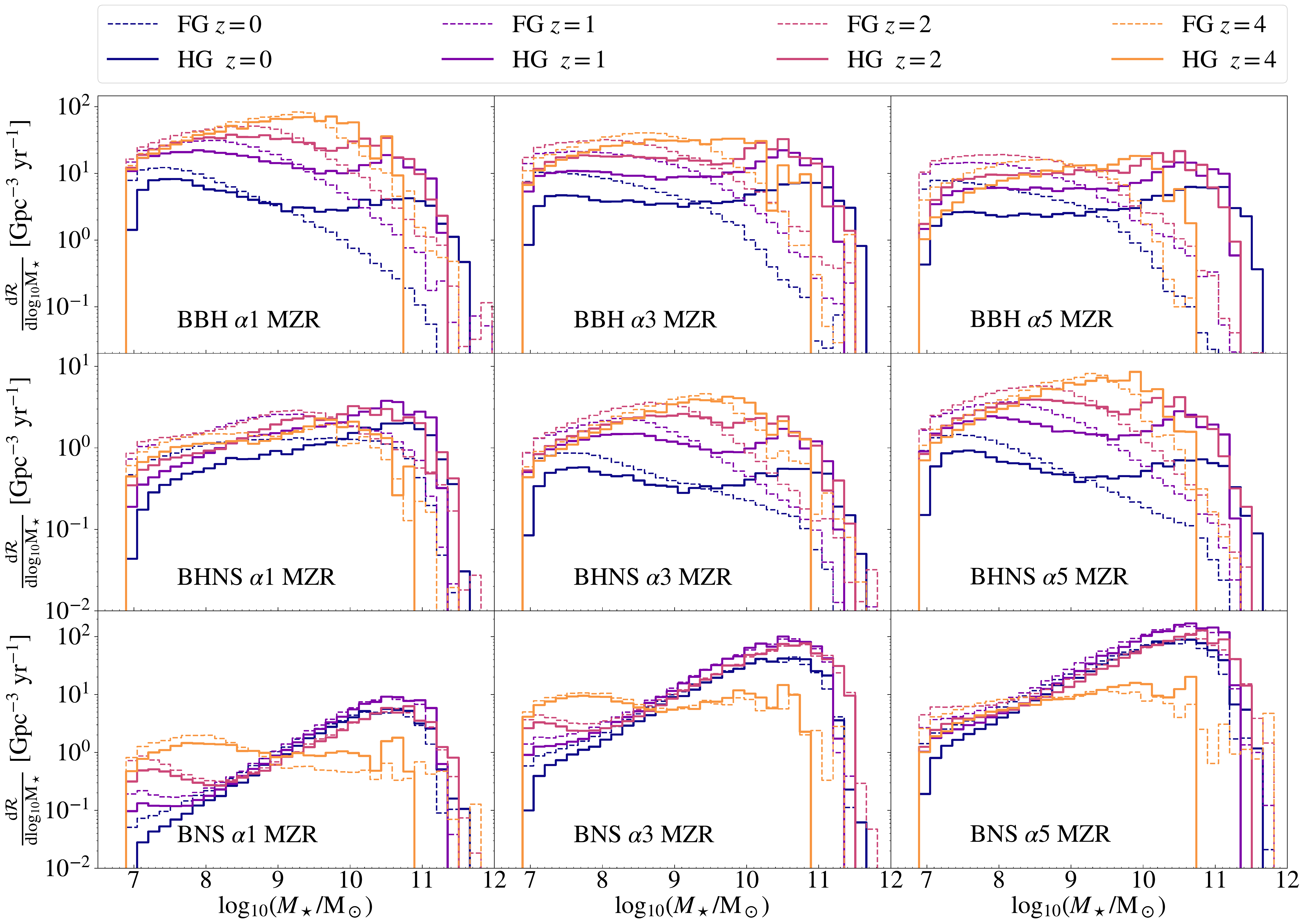}
    \caption{Distribution of the merger rate density for four different redshifts as a function of the formation galaxy (FG) mass (dashed lines) and host galaxy (HG) mass (solid lines), assuming the MZR. The different colours refer to redshift $z=0,$ 1, 2, and 4. 
    }
    \label{fig:mass_mzr}
\end{figure*}

\begin{figure*}
    \centering
    \includegraphics[width = 
    0.85\textwidth]{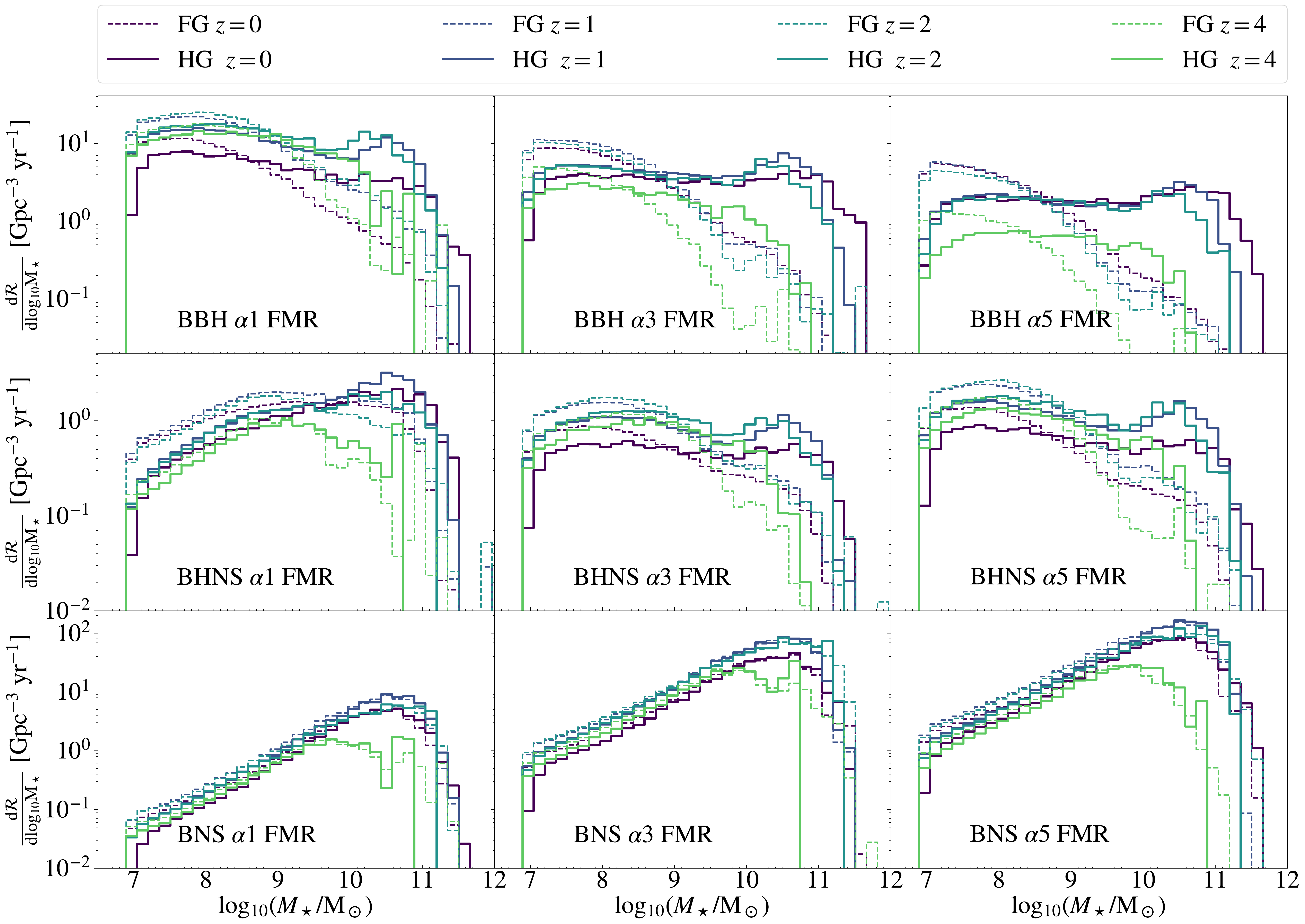}
    \caption{Same as Fig.~\ref{fig:mass_mzr}, but for the FMR.} 
    \label{fig:mass_fmr}
\end{figure*}

Figures~\ref{fig:mass_mzr} and \ref{fig:mass_fmr} show the mass distribution of FGs (dashed lines) and HGs (solid lines) 
for the MZR and the FMR, respectively. The time elapsed between compact-object formation and merger 
spans from a few Myr to several Gyr (Appendix \ref{sec:delay}).   
Hence, the HG might show significantly different properties with respect to the FG, not only because the original FG might have merged into a larger galaxy, but also because the FG might have drastically evolved (in terms of mass, SFR and metallicity) from the formation to the merger epoch of the BCO.

In the case of BNSs, the mass of the HG is always very similar to that of the  FG, because BNSs merge with a short delay time and are not affected by metallicity. In contrast, the FGs of BBHs and BHNSs tend to be less massive than the HGs. This 
happens because many FGs merge with other galaxies by the time of the BBH/BHNS  coalescence. 
The HGs in the local Universe are more massive than in any other previous epochs. 
 
 The fraction of high-mass HGs increases with increasing $\alpha$. In fact, $\alpha = 5$  corresponds to longer delay times on average. If the delay times are longer, the FGs have more time to merge with  other galaxies and form more massive HGs. 
On the other hand, both Figure \ref{fig:mass_mzr} and \ref{fig:mass_fmr} show that a large fraction of BBH 
HGs are low-mass galaxies. This is 
more important for $\alpha = 1$, which corresponds to shorter delay times. 
Hence, the HG mass distribution is shaped by the delay time distribution of compact objects, consistently with the result of previous authors \citep[e.g.,][]{mapelli2018,artale2019,artale2020a,mccarthy2020}. 

The HG mass distributions of both BBHs and BHNSs depend on the adopted metallicity model. At  $z \gtrsim 2$ the normalisation of the distribution is higher if we consider the MZR. This was also evident by looking at Figure \ref{fig:mrd}, where at $z \gtrsim 2$ the MZR yields many more BBH mergers than the FMR.

Finally, the HGs of BNSs tend to be more massive than the HGs of both BHNSs and BBHs, at least at $z<4$. The reason is that BNSs are insensitive to metallicity, while BHNSs and BBHs form predominantly in metal-poor (hence, smaller) galaxies. Both the FMR and the MZR predict that metal-poor galaxies are generally less massive than metal-rich ones.

\subsubsection{Star formation rate (SFR)}

\begin{figure*}
    \centering
    \includegraphics[width = 
    0.85\textwidth]{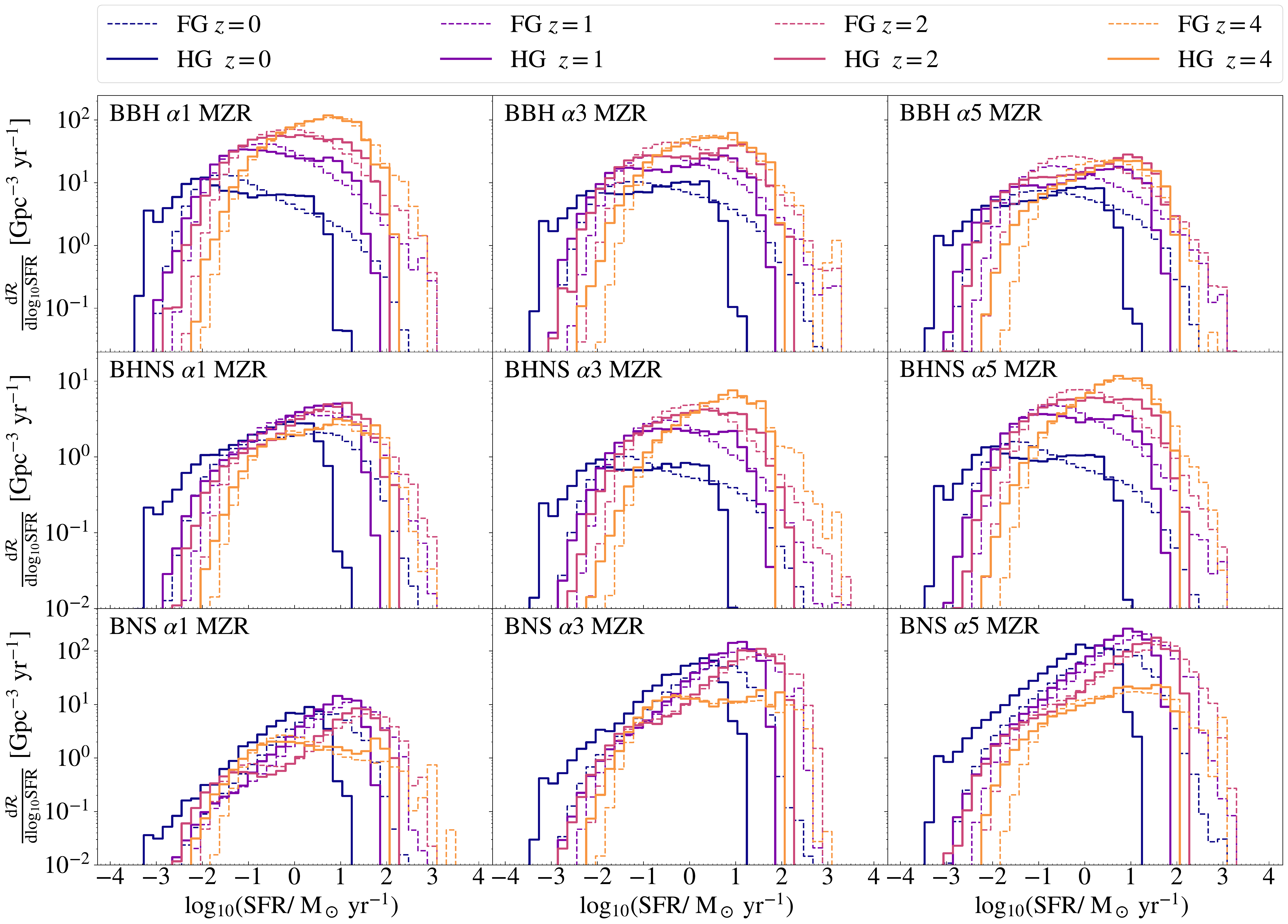}
    \caption{Distribution of the merger rate density for four different redshifts ($z=0,$ 1, 2 and 4) as a function of the SFR of the FGs (dashed lines) and HGs (solid lines), assuming the MZR.}
    \label{fig:sfr_mzr}
\end{figure*}
\begin{figure*}
    \centering
    \includegraphics[width=0.85\textwidth]{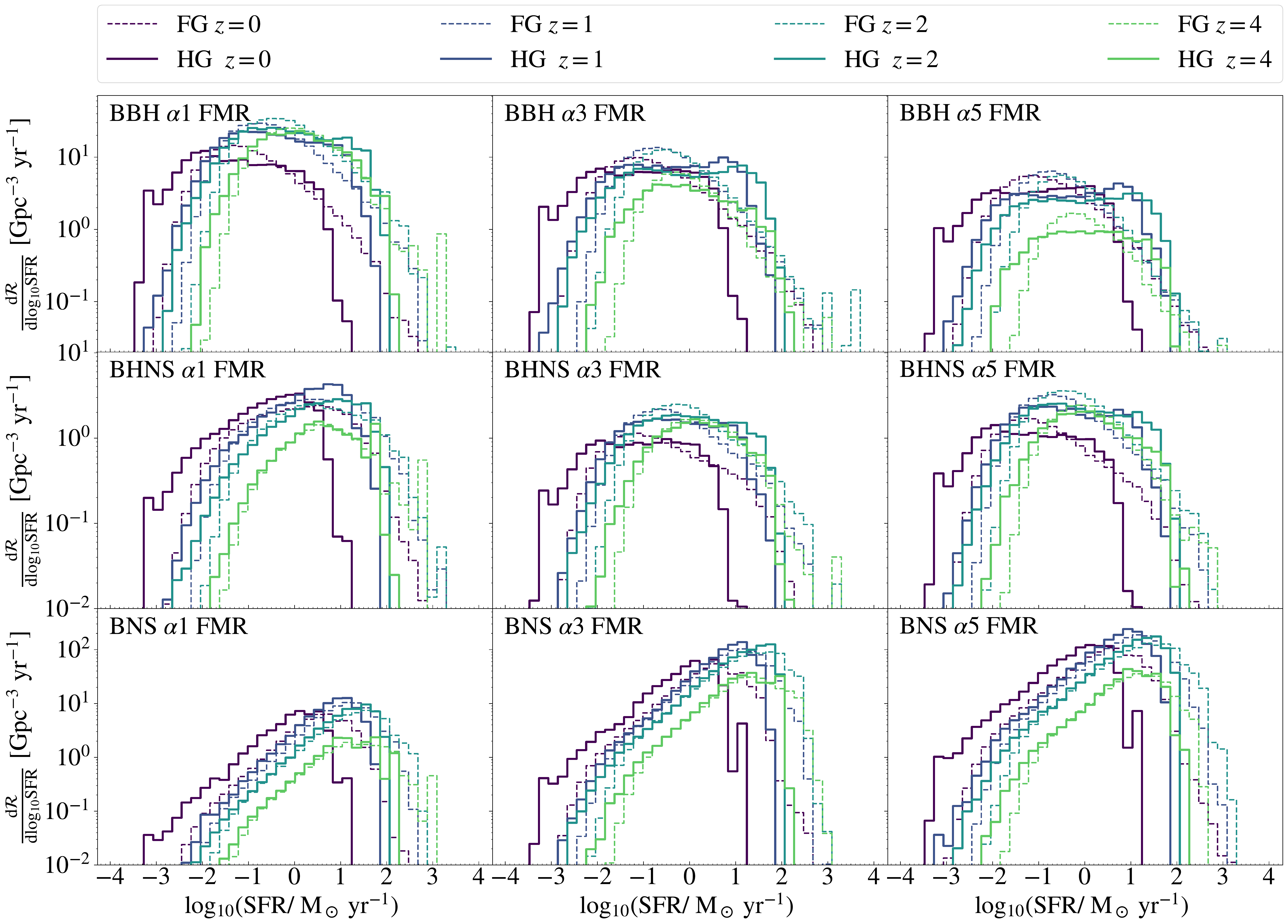}
    \caption{Same as Fig.~\ref{fig:sfr_mzr}, but for the FMR.} 
    \label{fig:sfr_fmr}
\end{figure*}

We show the SFR distribution of FGs and HGs in Figure   \ref{fig:sfr_mzr} for the MZR and Figure \ref{fig:sfr_fmr} for the FMR. 
At lower redshifts, compact objects tend to form and merge in galaxies with lower SFR.  The reason is that low-redshift galaxies have lower SFR on average, as shown in Figure~\ref{fig:sfrd}. 
At higher redshift ($z > 1$), the SFR distribution of FGs and HGs are very similar, since the delay times are short. 

At $z = 4$, the SFR of FGs and HGs of BBHs  peaks at lower values  when we assume the FMR than in the case of the MZR.
The reason is that the metallicity of massive and high star-forming galaxies is high enough to quench the formation of BBHs already at high redshift, if we assume the FMR.
The common envelope parameter $\alpha$ affects only the normalisation of the distributions.

\subsubsection{Metallicity}

\begin{figure*}
    \centering
    \includegraphics[width = 
    0.85\textwidth]{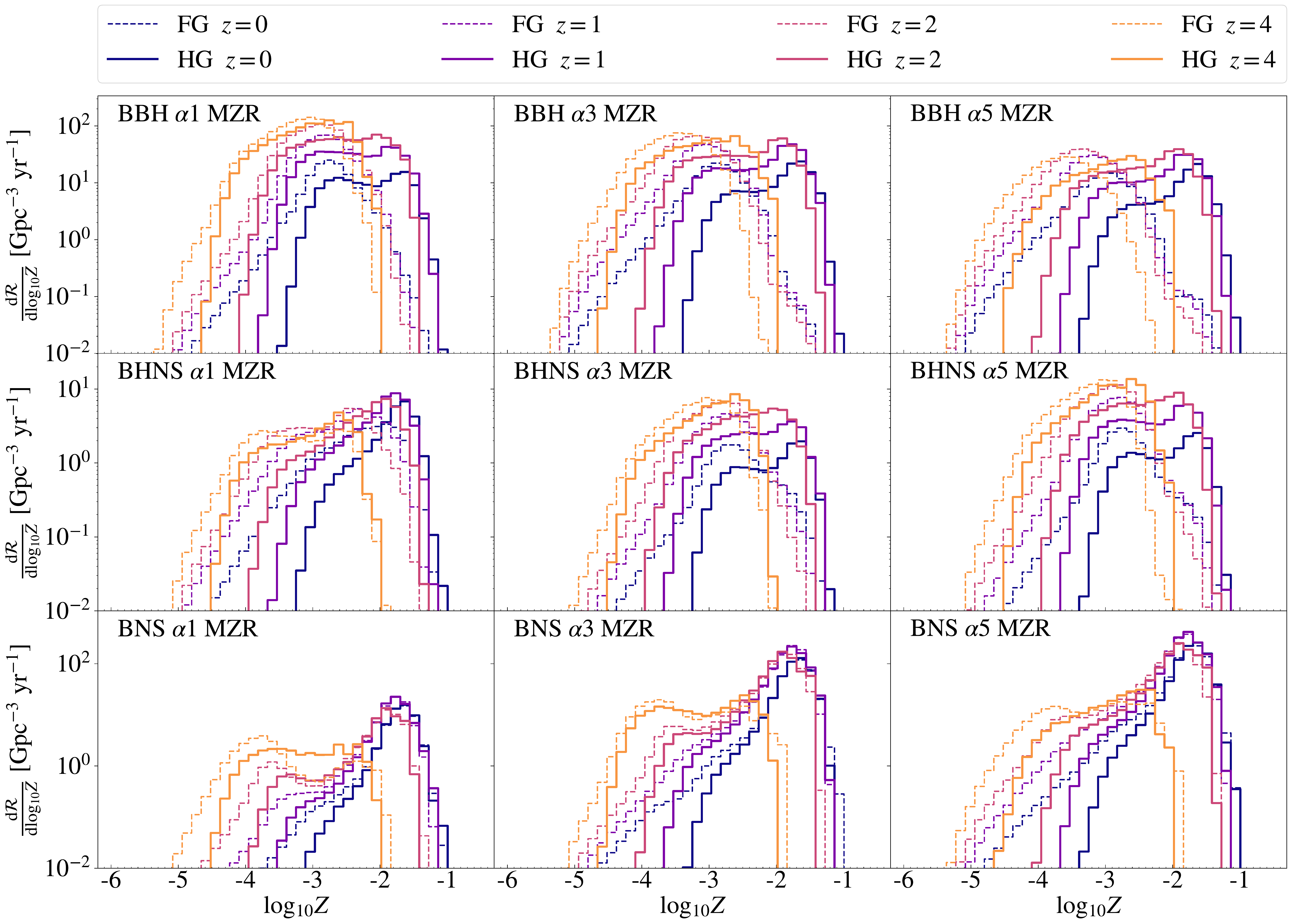}
    \caption{Distribution of the merger rate density for four different redshifts ($z=0,$ 1, 2, and 4) as a function of the FG metallicity (dashed lines) and HG metallicity (solid lines), assuming the MZR.}
    \label{fig:met_mzr}
\end{figure*}

\begin{figure*}
    \centering
    \includegraphics[width = 
    0.85\textwidth]{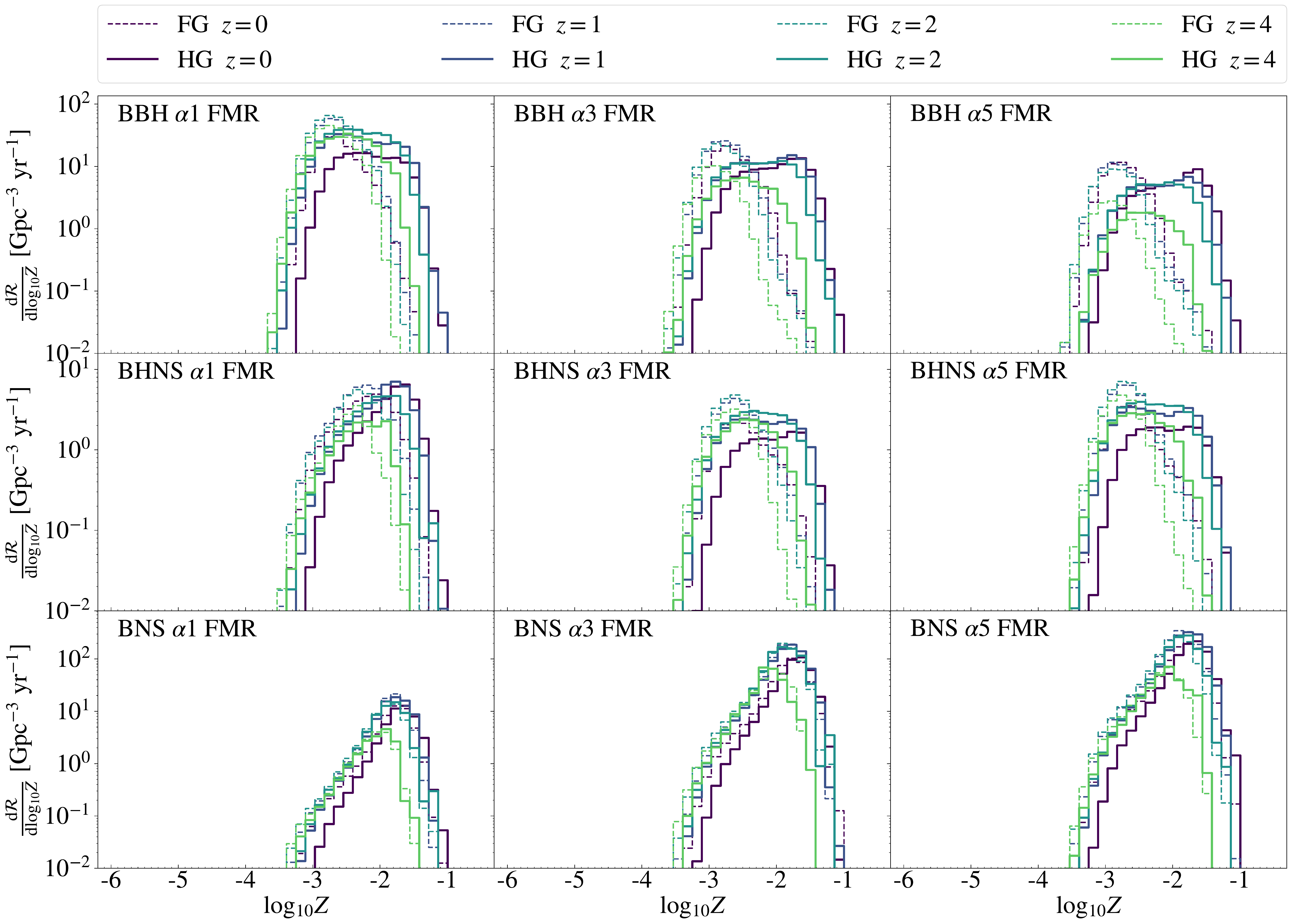}
    \caption{Same as Fig.~\ref{fig:met_mzr} but for the FMR.} 
    \label{fig:met_fmr}
\end{figure*}

Figure \ref{fig:met_mzr} (Figure \ref{fig:met_fmr}) shows the metallicity distribution of FGs and HGs we obtain if we adopt the MZR (FMR). 
The main difference between Figure~\ref{fig:met_mzr} and \ref{fig:met_fmr} is that 
 the distributions of both FGs and HGs extend to very low-metallicity tails in the case of the MZR, while they are  truncated at $\log_{10}{Z}\approx{-3.5}$ in the case of the FMR.

For both the MZR (Figure~\ref{fig:met_mzr}) and FMR (Figure~\ref{fig:met_fmr}), the metallicity of the HG tends to be higher than the metallicity of the FG, because the average metallicity of the Universe increases as redshift decreases (Figure \ref{fig:sfrd_zZ}). The metallicity difference between  HGs and  FGs is particularly large for BBHs and BHNSs, which merger efficiency depends on metallicity, while it is negligible for BNSs.

Overall, BBHs tend to form in metal-poor galaxies and merge in metal-rich galaxies. This is particularly evident in the local Universe. 
 In the case of BNSs, the peak of the metallicity distribution of FGs and HGs almost coincide. 


\subsection{Merger rate per galaxy} 
\label{sec:ngw}

\begin{figure*}
    \centering
    \includegraphics[width = 
    0.90\textwidth]{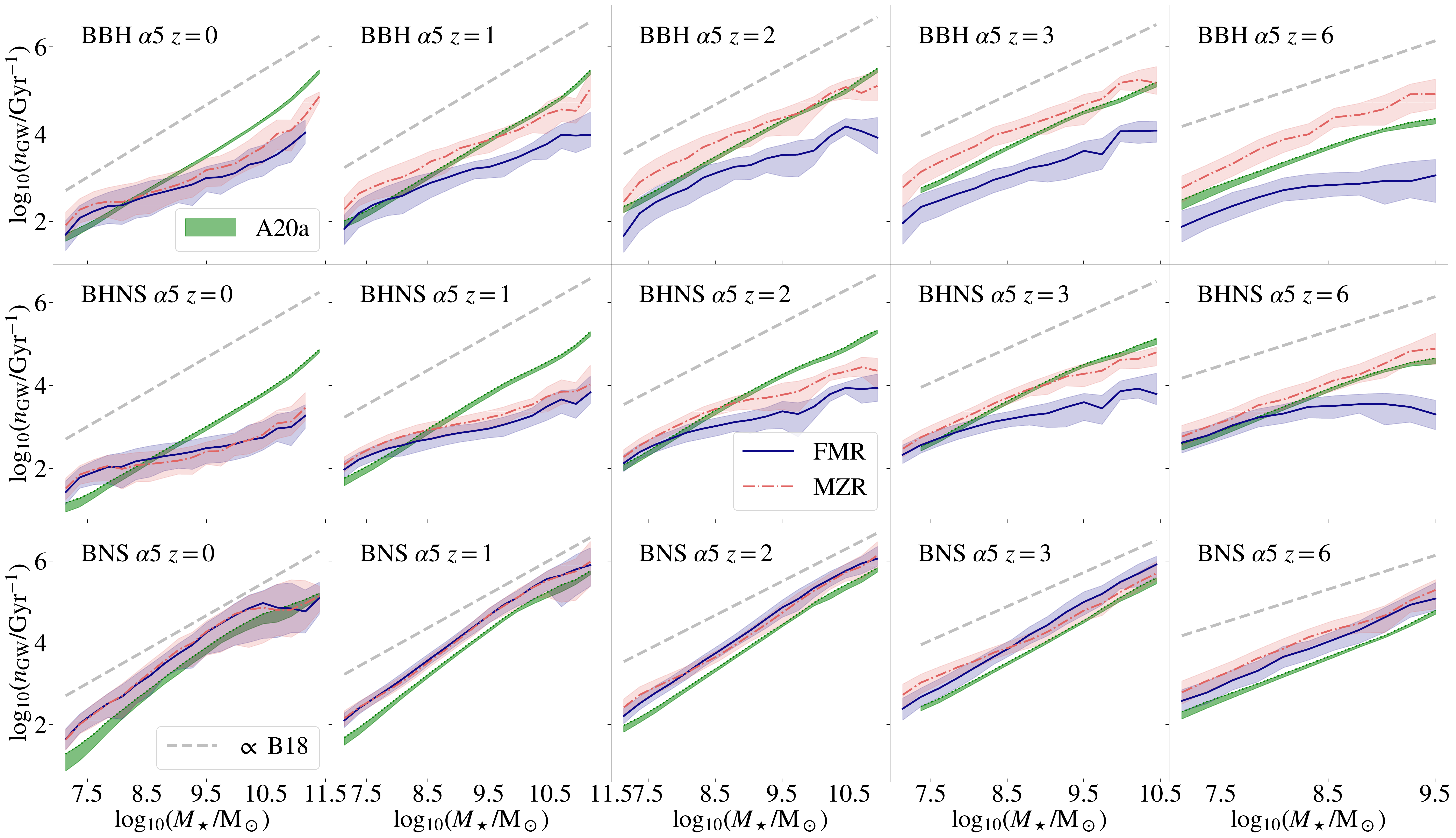}
    \caption{From left to right: median value of the merger rate per galaxy ($n_{\rm{GW}}$) as a function of the stellar mass $M_{*}$ for the HGs of BBHs (upper panels), BHNSs (middle panels) and BNSs (lower panels), for $z_{\rm{merg}} = 0$, 1, 2, 3 and  6. 
    The MZR and FMR are shown in red and blue, respectively. The shaded areas are the 75th and 25th percentiles of the distribution of $n_{\rm{GW}}$ for each stellar mass bin. The  dashed gray lines show the B18 main sequence definition with arbitrary normalisation. The green shaded areas are the results of \protect \citet[][A20a]{artale2020a}. }
    \label{fig:mr_z_mass}
\end{figure*}

\begin{table*}
    \centering
\resizebox{\textwidth}{!}{    \begin{tabular}{c|c|c|c|c|c|c|c|c|c|c|c}
    \toprule
    & & \multicolumn{2}{c|}{$z = 0$} &\multicolumn{2}{c|}{$z = 1.0$} & \multicolumn{2}{c|}{$z = 2.0$} & \multicolumn{2}{c|}{$z = 3.0$} &\multicolumn{2}{c|}{$z = 6.0$} \\
 
 &  & $a$ & $b$ & $a$ & $b$ & $a$ & $b$& $a$ & $b$ & $a$ & $b$ \\ 
 \hline

\multirow{3}{*}{BBH} & FMR & $-1.5 \pm 0.2$ & $0.47 \pm 0.02$ & $-1.5 \pm 0.2$ & $0.50 \pm 0.02$ & $-1.9 \pm 0.4 $ & $0.56 \pm 0.04  $ & $-2.2 \pm 0.2 $ & $0.61 \pm 0.03$ & $-1.0 \pm 0.5$ & $0.44 \pm 0.06$ \\
& MZR & $-2.3 \pm 0.3$ & $0.58 \pm 0.03$ & $-1.8 \pm 0.2$ & $0.60 \pm 0.02$ & $-1.9 \pm 0.3$ & $0.66 \pm 0.03$ & $-2.2 \pm 0.2$ & $0.72 \pm 0.03$ & $-3.7 \pm 0.4$ & $0.93 \pm 0.04$ \\
& A20a & $-4.5 \pm 0.2$ & $0.85 \pm 0.02$ & $-4.1 \pm 0.1$ & $0.84 \pm 0.01$ & $-3.56 \pm 0.05$ & $0.83 \pm 0.01$ & $-3.1 \pm 0.1$ & $0.79 \pm 0.01$ & $-3.2 \pm 0.2$ & $0.81 \pm 0.02 $ \\

\multirow{3}{*}{BHNS} & FMR & $-0.9 \pm 0.2$ & $0.36 \pm 0.02$ & $-0.8 \pm 0.2$ & $0.40 \pm 0.02$ & $-0.8 \pm 0.2$ & $0.45 \pm 0.02$ & $-0.60 \pm 0.2$ & $0.44 \pm 0.03$ & $0.54 \pm 0.66$ & $0.33 \pm 0.08$ \\
& MZR & $-1.1 \pm 0.3$ & $0.38 \pm 0.03$ & $-0.8 \pm 0.1$ & $0.43 \pm 0.01$ & $-1.3 \pm 0.2$ & $0.53 \pm 0.03$ & $-2.2 \pm 0.2$ & $0.68 \pm 0.03$ & $-3.7 \pm 0.2$ & $0.91 \pm 0.02$ \\
& A20a & $-5.0 \pm 0.1$ & $0.85 \pm 0.01$ & $-4.41 \pm 0.07$ & $0.86 \pm 0.01$ & $-3.9 \pm 0.1$ & $0.85 \pm 0.01$ & $-3.7 \pm 0.2$ & $0.86 \pm 0.02$ & $-3.9 \pm 0.2$ & $0.92 \pm 0.02$ \\

\multirow{3}{*}{BNS} & FMR & $-3.9 \pm 0.5$ & $0.82 \pm 0.05$ & $-4.8 \pm 0.2$ & $0.98 \pm 0.02$ & $-5.1 \pm 0.1$ & $1.04 \pm 0.02$ & $-5.29 \pm 0.07$ & $1.08 \pm 0.01$ & $-5.1 \pm 0.1$ & $1.08 \pm 0.02$ \\

& MZR & $-3.9 \pm 0.4$ & $0.83 \pm 0.05$ & $-4.9 \pm 0.2$ & $0.99 \pm 0.02$ & $-4.7 \pm 0.1$ & $0.99 \pm 0.01$ & $-3.5 \pm 0.2$ & $0.87 \pm 0.02$ & $-4.4 \pm 0.2$ & $1.02 \pm 0.02$ \\

& A20a & $-5.3 \pm 0.3$ & $0.95 \pm 0.03$ & $-5.6 \pm 0.2$ & $1.04 \pm 0.02$ & $-5.53 \pm 0.07$ & $1.05 \pm 0.01$ & $-5.27 \pm 0.09$ & $1.04 \pm 0.01$ & $-5.0 \pm 0.1$ & $1.02 \pm 0.02$ \\

\bottomrule
    \end{tabular}}
    \caption{Fits of the merger rate per galaxy $\log_{10} (n_{\rm{GW}}/ {\rm{Gyr}}^{-1}) = a + b \log_{10}( M_*/{\rm{M}}_\odot)$ at $z = 0, 1, 2, 3$ and $6$ for the models shown in Figure \protect \ref{fig:mr_z_mass}.} 
    \label{tab:ngw_fits}
\end{table*}

Figure \ref{fig:mr_z_mass} shows the merger rate per galaxy $n_{\rm{GW}}$ as function of the stellar mass. We compare our results to those of \cite{artale2019} and \cite{artale2020a}, who adopt the {\sc eagle} cosmological simulation to retrieve this information. The merger rate of BNSs strongly correlates with the stellar mass of the HG. The correlation slope (Table \ref{tab:ngw_fits}) is consistent with that obtained by \cite{artale2020a} with cosmological simulations in the case of BNSs. 
To interpret this result, in Figure \ref{fig:mr_z_mass} we show the main sequence definition by B18, with an arbitrary normalisation. It is apparent that the correlation of $n_{\rm{GW}}$ with the stellar mass is mainly a consequence of the main sequence of star-forming galaxies.  

Deviations from the main sequence are particularly evident in the case of BBHs and BHNSs. In fact, the merger rate per galaxy of BBHs and BHNSs also depends on the chosen metallicity relation. The merger rates per galaxy obtained with the MZR or FMR are almost identical in the local Universe, but become dramatically different at redshift $z\gtrsim{1}$, in terms of both slope and normalization.  
This happens because the differences between FMR and MZR increase with redshift (Figure \ref{fig:met_comp}). 
In fact, in the case of the FMR the formation of BBHs and BHNSs drops in galaxies with  $M_*>10^9$ M$_\odot$ at $z \geq 4$ (Figure \ref{fig:mass_fmr}).  As a result, the BBH merger rate per galaxy at $z=6$ has a much shallower trend with $M_*$ in the case of the FMR with respect to the MZR.

\subsection{Role of passive galaxies}

\begin{figure*}
    \centering
    \includegraphics[width = 
    0.80\textwidth]{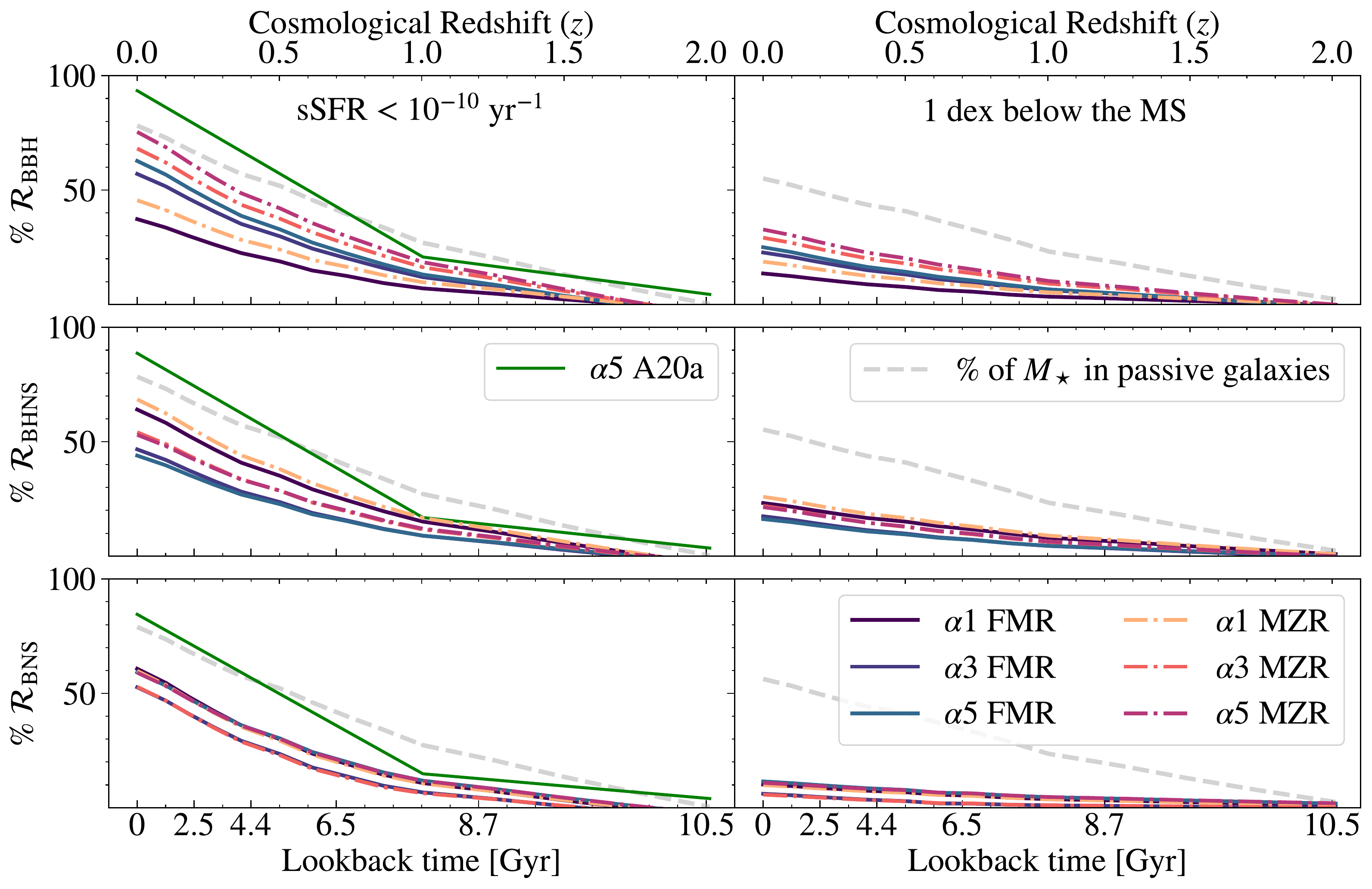}
    \caption{Percentage of compact object mergers 
    in passive galaxies for the models considered in this work. 
    Upper, middle and lower panel: BBHs, BHNSs and BNSs. 
    Left-hand column: definition of passive galaxies from  \protect\citet[][A20a]{artale2020a}, i.e. passive galaxies have $\rm{sSFR}<10^{-10}$ yr$^{-1}$. Green solid line: results from A20a. Right-hand column: our definition of passive galaxies (i.e., galaxies with SFR at least 1 dex below the star-forming main sequence, Section \protect\ref{sec:passive}). The grey dashed line is the percentage of stellar mass ($M_*$) stored in passive galaxies as function of redshift, according to the two above-mentioned definitions.}  
    \label{fig:perc_pass} 
\end{figure*}

Figure \ref{fig:perc_pass} shows the percentage of compact object mergers in passive galaxies. 
Besides our definition of passive galaxies (Section~\ref{sec:passive}), 
we also show the definition adopted by \cite{artale2019}, for comparison. 
According to the definition in \cite{artale2019}, passive galaxies are galaxies with specific SFR (sSFR) $< 10^{-10}$ yr$^{-1}$. Figure \ref{fig:perc_pass} shows that the contribution of passive galaxies to the merger rate increases as redshift decreases, for both  definitions. 
This happens because the  percentage of stellar mass stored in passive galaxies 
increases as we approach the local Universe. 

With common envelope efficiency  $\alpha=5$, which corresponds  to longer delay times, the fraction of BBH mergers in passive galaxies is higher with respect to lower values of $\alpha$. 
The MZR predicts more BBH mergers in passive galaxies at fixed redshift with respect to the FMR, while the fraction of BNS mergers in passive galaxies does not depend on  
the chosen metallicity evolution. 

Figure \ref{fig:perc_pass} also shows the impact of different definitions of passive galaxies: with the definition adopted in this work, we estimate less than a half (a fifth) of BBH (BNS) mergers in passive galaxies with respect to \cite{artale2020a}.

\section{Discussion}
\label{sec:discussion}

Here, we discuss some of the main assumptions we made in our model and their impact on our results. We will not consider the impact of binary population synthesis parameters, which we have already described in \cite{santoliquido2021}. 

\subsection{Constant versus variable GSMF}

\begin{figure}
    \centering
    \includegraphics[width = 0.45 \textwidth]{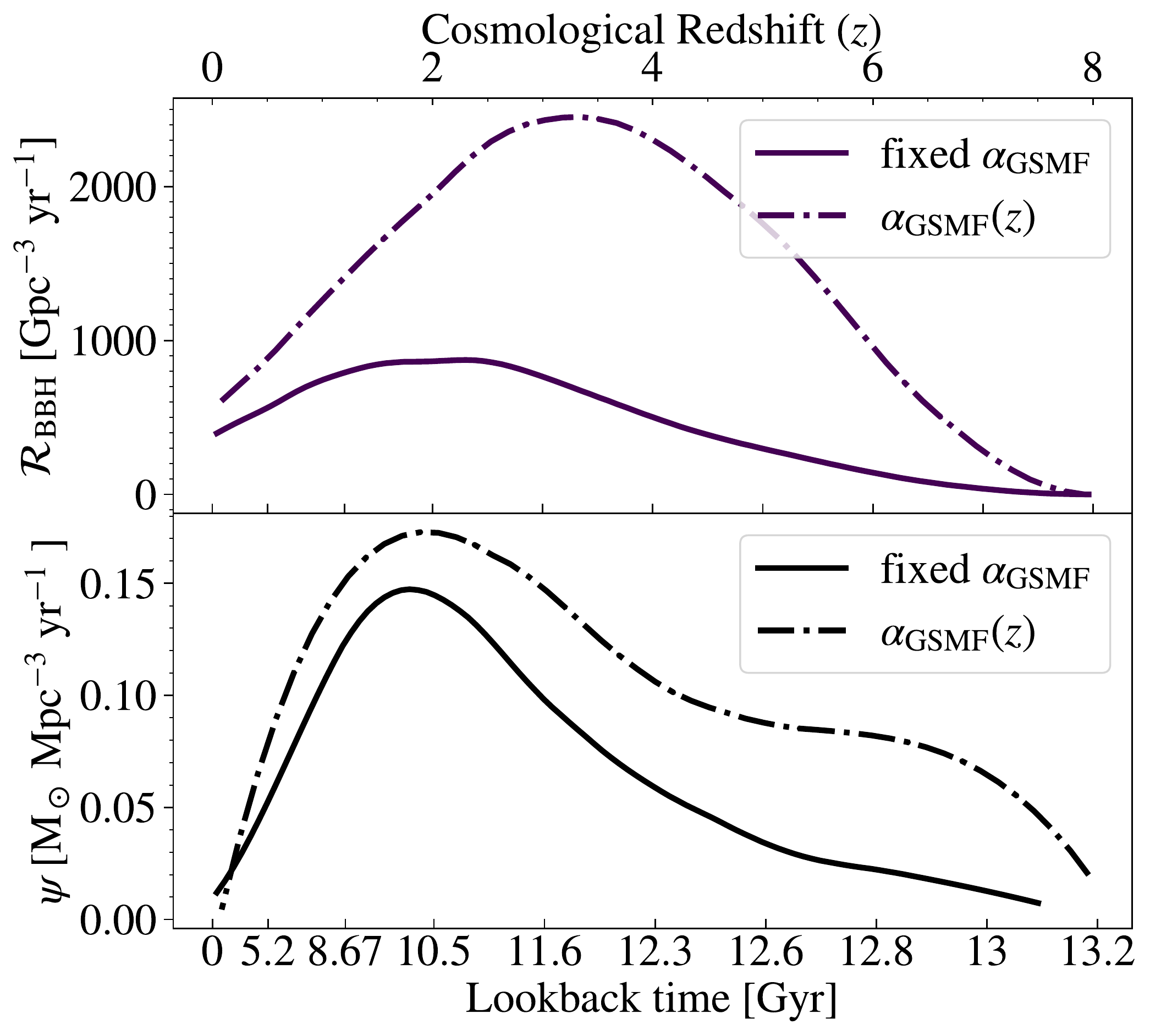}
    \caption{Upper panel: Evolution of the BBH merger rate density $\mathcal{R}_{\rm BBH}(z)$ calculated with a constant value of $\alpha_{\rm GSMF}$ (solid line) and with the value $\alpha_{\rm GSMF}(z)$ proposed by \protect\cite{chruslinska2019} (dot-dashed line). In both cases, we assume $\alpha=5$, MZR, S14 and $M_{\rm{min}} = 10^7 ~{\rm{M}}_\odot$. Lower panel: cosmic SFR density $\psi(z)$ obtained adopting the two different GSMFs shown in the upper panel.} 
    \label{fig:gsmf_var}
\end{figure}

We  assumed that the slope of the GSMF ($\alpha_{\rm{GSMF}}$)  is constant with redshift. \cite{chruslinska2019} pointed out that the slope of the low-mass end of the GSMF is weakly constrained. Although the low-mass galaxies are overall the most abundant, they are also the faintest and most difficult to observe, especially at high redshift. Table 1 in  \cite{chruslinska2019} shows that the low-mass slope tends to steepen with redshift. In other words, $\alpha_{\rm{GSMF}}$ in Equation \ref{eq:gsmf} becomes more negative at increasing redshift (Figure 3 of \citealt{chruslinska2019}). 
 Figure \ref{fig:gsmf_var} shows the impact of the evolving low-mass slope of GSMF with redshift, $\alpha_{\rm{GSMF}}(z)$, on the cosmic SFR density and on the merger rate density. For the latter, we  show the case of BBHs, $\alpha=5$ and MZR to maximise the impact of a higher fraction of metal-poor low-mass galaxies on the merger rate density. The cosmic SFRD varies at most by a factor of 4 at $z\sim 6$, resulting in a similar difference 
 at $z\sim3.5$ in the merger rate density.

\subsection{Main sequence of star forming and starburst galaxies}

\begin{figure}
    \centering
    \includegraphics[width = 0.45 \textwidth]{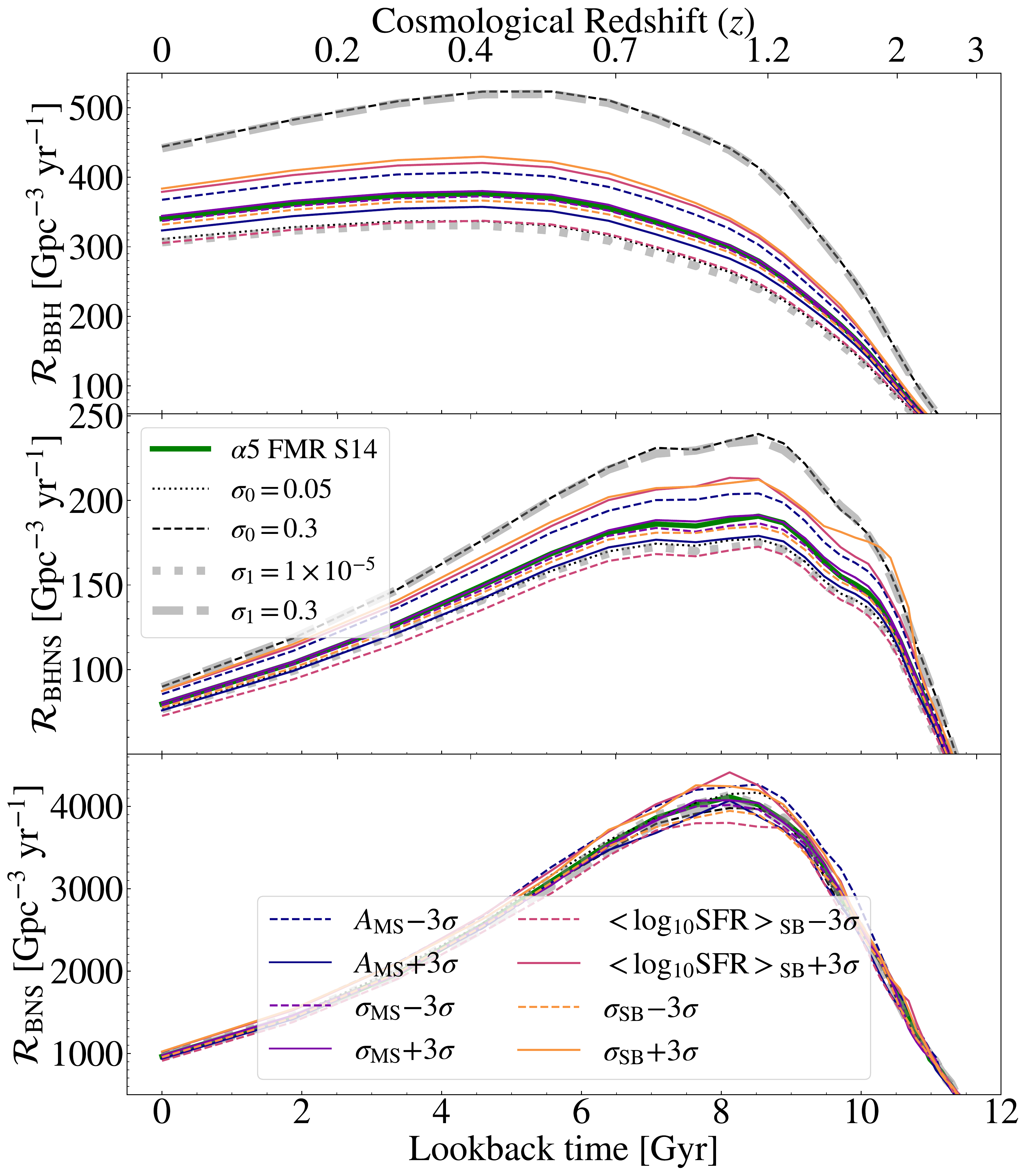}
    \caption{
    Impact of $\pm 3 \sigma$ variation of the parameters in Equation \protect \ref{eq:2gauss} on the merger rate density $\mathcal{R}(z)$ of BBHs (upper panel), BHNSs (middle panel), and BNSs (lower panel). We assume, as reference, $\alpha=5$, FMR, S14 and $M_{\rm{min}} = 10^7 ~{\rm{M}}_\odot$ with $\sigma_0 = 0.15$ and $\sigma_1 = 0.14$ (thin green line). Solid and dashed dark-blue lines: $A_{\rm{MS}}\pm{}3\,{}\sigma$; solid and dashed violet lines:  $\sigma_{\rm{MS}}\pm{}3\,{}\sigma$; solid and dashed pink lines: $\langle{}\log_{10}{\rm{SFR}}\rangle{}_{\rm{SB}}\pm{}3\,{}\sigma$; solid and dashed orange lines: $\sigma_{\rm{SB}}\pm{}3\,{}\sigma$; thin dotted and dashed black lines: $\sigma_0=0.05$ and 0.3; thick dotted and dashed grey lines: $\sigma_1=10^{-5}$  and 0.3} 
    \label{fig:referee}
\end{figure}

We assumed that the star-forming main sequence has no flattening at high masses, i.e. the linear relation is preserved at any value of stellar mass (equations  \ref{eq:ms_speagle} and \ref{eq:boogaard}).  \cite{chruslinska2019} show that there can be two 
main variations with respect to this assumption. The first variation is referred to as \textit{moderate flattening}. In this case the high-mass end of the main sequence is less steep than that of the low-mass end, and can also evolve with redshift becoming steeper with increasing $z$ \citep{speagle2014, boogaard2018, popesso2019}. The second variation is called \textit{sharp flattening} and the main sequence has an 
even  sharper flattening at high masses \citep{tomczak2016}. The resulting SFR is almost constant with increasing stellar mass (Figure 5 of \citealt{chruslinska2019}).  
We expect that the impact on our results of the moderate and sharp flattening of the main sequence of massive galaxies is lower with respect to changing the definition of the main sequence itself (i.e., considering both S14 and B18).

The distribution of star-forming galaxies at fixed mass (Equation \ref{eq:2gauss}) also relies on a number of parameters affected by observational uncertainties: $A_{\rm MS}$, $A_{\rm SB}$, $\sigma_{\rm MS}$, $\sigma_{\rm SB}$, $\langle{}\log_{10}{\rm SFR}\rangle{}_{\rm MS}$ and $\langle{}\log_{10}{\rm SFR}\rangle{}_{\rm SB}$ (see Table~\ref{tab:param} and Section~\ref{sec:sfr} for details). Figure \ref{fig:referee} shows the impact of $\pm3\sigma$ variations of these parameters on the merger rate density. The shift between starburst galaxies and main sequence (i.e., the definition of  $\langle{}\log_{10}{\rm SFR}\rangle{}_{\rm SB}$) yields the largest differences ($\pm 10\%$) on the merger rate density of BBHs at $z = 0$. 
Overall, BNSs are less affected by these parameters than both BBHs and BHNSs, because they are less sensitive to metal-dependent star formation.

Recent papers (e.g. \citealt{caputi2017} and \citealt{bisigello2018}) show  that the percentage of starburst galaxies with respect to all star-forming galaxies 
might increase towards lower stellar masses and with redshift. This suggests that the contribution of starburst galaxies to the total cosmic SFR density is higher than our main assumption (Section \ref{sec:sfr}). 
To see an example of the impact of this treatment of starburst galaxies on the SFRD$(z,Z)$, we refer to \cite{chruslinska2021} (Figure 10). We will include the impact of this treatment of starburst galaxies in a follow-up study.

\subsection{Metallicity relationships}

The intrinsic scatter around both the MZR and the FMR is also uncertain, as well as the metallicity gradient within each galaxy, described by $\sigma_0$ and $\sigma_1$, respectively (Section \ref{sec:metallicity}). We varied these parameters to assess their impact on the merger rate density, as shown in Figure \ref{fig:referee}. In the local Universe,  the merger rate density of BBHs varies at most by $-5\,{} (+15)$~\% and $-5\,{} (+19)$~\% if we consider $\sigma_0 ~(\sigma_1) = 0.05 ~(1\times10^{-5}) $ and 0.30 (0.30), respectively.

Previous works have proposed a third observational relation to describe the metallicity evolution of galaxies: 
the fundamental plane  \citep{laralopez2010, hunt2012,hunt2016}. 
Similarly to the FMR, the fundamental plane is independent of redshift, 
and relates metallicity, SFR and stellar mass. However, 
in the case of the fundamental plane these three quantities are linked in a two-dimensional plane. In this way, the value of the galaxy metallicity can be expressed as \citep{hunt2016}:  
\begin{equation}
    \label{eq:fp}
    12 + \log_{10} ({\rm O/H}) = -0.14\,{}\log_{10} {\rm{SFR}} + 0.37\,{}\log_{10}{M_*} + 4.82.
\end{equation}
As a result, at high mass there is no bending after the turn-over mass and thus the metallicity does not converge to an asymptotic value. At low mass instead, Equation \ref{eq:fp} yields an even 
flatter trend with stellar mass at fixed SFR with respect to the FMR. Thanks to this behaviour, the fundamental plane yields a merger rate density evolution in between the FMR and the MZR.

\begin{figure}
    \centering
    \includegraphics[width = 0.48 \textwidth]{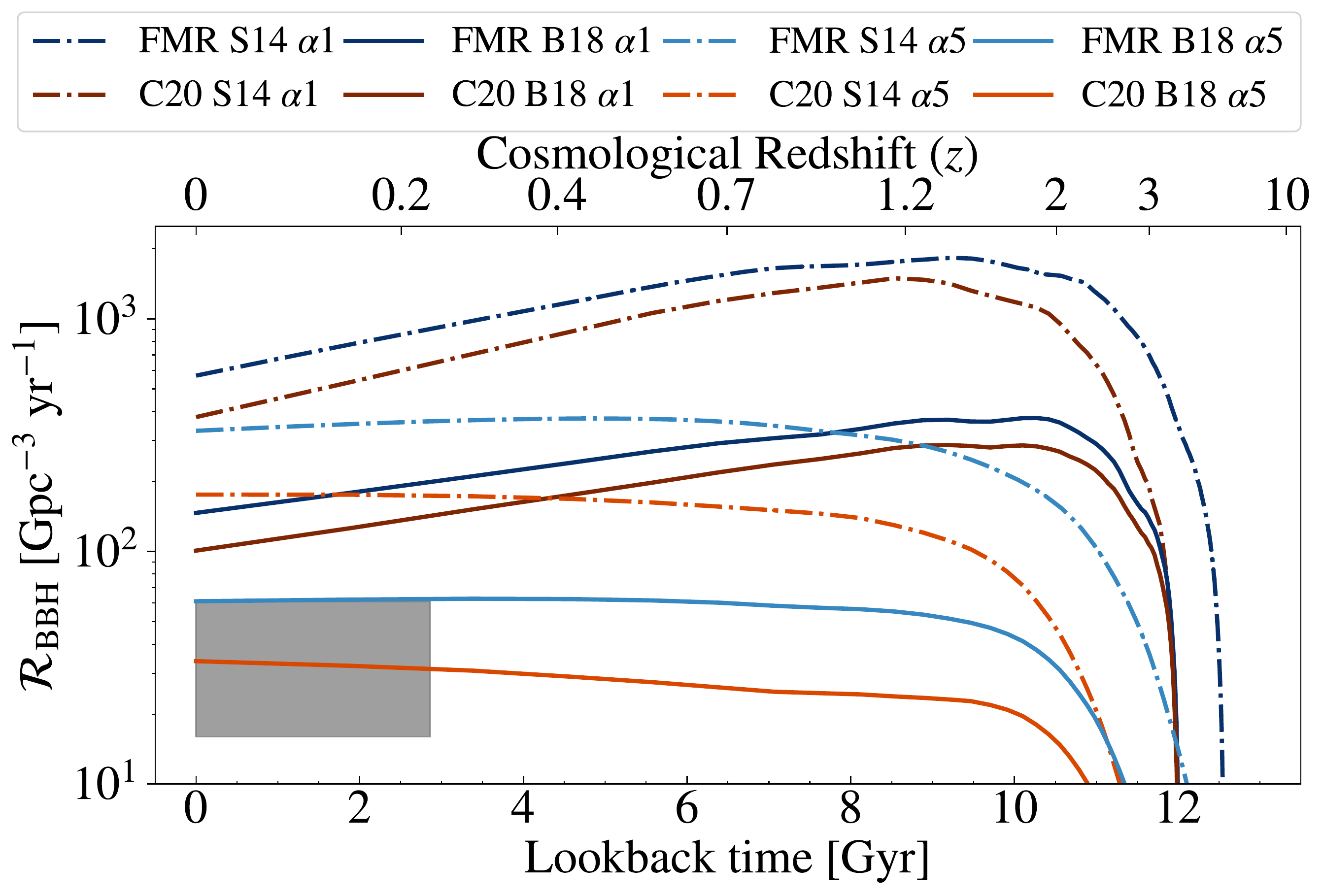}
    \caption{
    Evolution of the BBH merger rate density $\mathcal{R}_{\rm BBH}(z)$ obtained with the fundamental metallicity relation defined in \protect \citet[][C20]{curti2020}  and \protect \citet[][FMR]{mannucci2011} 
    for $\alpha=1$ and $5$. 
    Solid and dash-dotted lines refer to B18 and S14, respectively. The gray shaded area shows the 90\% credible interval of the local BBH merger rate density \protect \citep{pop3_2021}.}
    \label{fig:mrd_curti}
\end{figure}

One last caveat concerns the calibration of metallicity in the adopted scaling relations.  
Empirical metallicity calibrations are one of the main sources of uncertainty in determining the metallicity of galaxies \citep{Kewley2008, maiolino2019, curti2020}. In fact, different metallicity calibrations can give rise to different a normalisation (up to $\sim 0.6$ dex) and shape of the MZR (see, for example, Figure 4 of \citealt{chruslinska2019}). To assess the impact of a different metallicity calibration  on our results, we evaluated the BBH merger rate density with the new definition of FMR 
given in Equation 5 of \citet[][hereafter C20]{curti2020}. The main difference between the FMR we adopted in our paper and the FMR derived by C20 is that the latter has a shallower slope for galaxies less massive than the turnover mass ($M_0 = 10^{10.02 \pm 0.09}\,{}{\rm{M}_\odot}$); in other words, galaxies with $M\lesssim 10^8$ M$_\odot$ have higher metallicity at any SFR (lower panel of Figure 3 of C20). Figure \ref{fig:mrd_curti} shows that the overall evolution 
of the BBH merger rate density with redshift is not heavily affected by this different metallicity calibration. On the other hand, the BBH merger rate density in the local Universe resulting from the FMR reported by C20 is a factor $\sim 1.2 - 2$ lower with respect to the FMR adopted in this work.

\section{Conclusions}
\label{sec:conlusion}
 
 We developed the new 
 code \galaxyrate{}, which estimates the merger rate density of binary compact objects (BCOs) and the properties of their host galaxies (HGs), based on observational scaling relations. 
With \galaxyrate{}, we can perform many realizations of the merger rate density and HG properties across the cosmic time, in order to bracket the main uncertainties springing from both BCO formation and galaxy evolution.

Here, we have adopted three BCO catalogues generated with the population-synthesis code {\sc mobse} \citep{giacobbo2018a}. In each catalogue, we vary the common-envelope efficiency parameter $\alpha=1$, 3 and 5. This parameter is 
one of the main sources of uncertainty among binary evolution processes \citep{santoliquido2021} and has a great impact on the delay time distribution of BCOs (Appendix \ref{sec:delay}). 

\galaxyrate{} first generates the formation galaxy (FG) of each BCO, i.e. the galaxy in which the progenitor stars of the BCO form. 
Each FG is described by its stellar mass, SFR and metallicity. We extract these properties from observational relations: the galaxy stellar mass function, the SFR  distribution of star-forming and starburst galaxies, the mass--metallicity relation (MZR), and the fundamental metallicity relation (FMR). Given the large observational uncertainties, we explored the parameter space (Section \ref{sec:galaxymodel}) that mostly affects our results. For instance, we considered two galaxy main sequence definitions (S14 and B18). We also compared the metallicity evolution obtained with the MZR and FMR. 

The HG (i.e., the galaxy in which the BCO merges) can be different from the FG for several reasons: the FG might have merged with other galaxies by the time the BCO reaches coalescence, or it might have evolved changing its mass, SFR and metallicity. We thus assign a HG to each BCO based on the conditional probability $\mathcal{P}(M_{\rm host}, {\rm{SFR}}_{\rm host}| M_{ \rm form}, {\rm{SFR}}_{\rm form}, z_{\rm form}, z_{\rm merg})$, i.e. the probability that the HG  has mass $M_{\rm host}$ and star formation rate ${\rm{SFR}}_{\rm host}$, given the mass ($M_{ \rm form}$) and SFR (${\rm{SFR}}_{\rm form}$) of the FG, and given the formation ($z_{\rm form}$) and merger redshift ($z_{\rm merg}$) of the BCO (Section \ref{sec:mergergalaxies}). 
Here, we calculate this probability in an empirical way by using the merger trees from the {\sc eagle} cosmological simulation 
\citep{schaye2015,Qu2017}.  

We found that the merger rate density evolution with redshift changes dramatically depending on the choice of the star-forming galaxy main sequence, especially in the case of BBHs and BHNSs. The local merger rate density of BBHs and BHNSs is $\sim{3-4}$ times higher if we assume the star-forming main sequence from S14 with respect to B18. This happens because the S14 main sequence predicts a significantly higher SFR density at high redshift with respect to B18 (Figure \ref{fig:sfrd}). BBHs and BHNSs are strongly affected by this difference, because their merger rate depends on metallicity and newly born stars at high redshift are preferentially metal-poor. In contrast, BNSs are marginally affected, because their merger rate does not depend on metallicity (Figure \ref{fig:mrd}).

The choice of the metallicity evolution has an important effect on the slope of the merger rate density of BBHs and BHNSs. The slope of the merger rate density evolution of BBHs and BHNSs is steeper if we assume the MZR with respect to the  FMR, because the latter predicts a shallower decrease of metallicity with redshift  (Figure \ref{fig:mrd}). In contrast, BNSs are not affected by the choice of the metallicity relation.

Also, we compared the merger rate density  obtained with \galaxyrate{}  (Section \ref{sec:MRD}) with that obtained with \cosmorate{} (Appendix \ref{sec:cosmorate}), which evaluates the merger rate density by assuming the average SFR density and metallicity evolution of the Universe, i.e. without information about the galaxies  \citep{santoliquido2021}. We found that the merger rate density evolution of BNSs obtained with \cosmorate{} and \galaxyrate{} are in good agreement, while the  BBH and BHNS merger rate densities evaluated with \galaxyrate{} are higher 
than those obtained with \cosmorate{}, if we assume the fitting formulas from \cite{madau2017} and a metallicity spread $\sigma_{\rm Z}=0.3$ (Figure~\ref{fig:mrd}). The main reason is that the SFRD$(z,Z)$ we obtain from the observational scaling relations supports a larger population of metal-poor stars with respect to the \cite{madau2017} fitting formulas with $\sigma_{\rm Z}=0.3$. The differences between the BBH merger rate density obtained with \galaxyrate{} and \cosmorate{} can be reconciled by assuming a larger metallicity spread $\sigma_{\rm Z} > 0.4$ in \cosmorate{}.  

The merger rate density of BNSs and BHNSs is within the 90\% credible interval estimated by the LVK with GWTC-3 \citep{pop3_2021} for all considered assumptions, while the merger rate density of BBHs predicted by our models is higher than the LVK range. This discrepancy most likely originates from the interplay of several different sources of uncertainty. Firstly, we considered only the formation of BBHs from binary evolution, and neglected the dynamical formation channel, which is  more effective for BBHs than for the other families of BCOs. Current models of binary evolution predict an extremely strong dependence of the BBH merger rate density on metallicity, while this dependence is quenched by most dynamical formation channels 
 \citep[e.g.,][]{mapelli2022}. It might be that the dependence of the BBH merger rate on metallicity is overestimated by current binary evolution models, or that most BBHs do not form via this channel. In a follow-up study, we will consider alternative formation channels for BBHs. Secondly, current models of binary evolution might 
 overestimate the BBH merger rate because of our poor knowledge of  several binary evolution processes, such as common envelope and the stability of Roche lobe \citep{marchant2021,klencki2021,gallos_garcia2021,belczynski2022}. Thirdly, the evolution of stellar metallicity across cosmic time is one of the most disputed aspects of galaxy evolution, because all the metallicity calibrations are affected by large (and sometimes systematic) uncertainties \citep[][]{maiolino2019}.


Overall, the HGs of BBHs and BHNSs are more massive than their FGs (both assuming the MZR and the FMR), because both BBHs and BHNSs tend to form in smaller metal-poor galaxies and to merge in larger metal-rich galaxies. In contrast, the FGs and HGs of BNSs are very similar to each other (Figures~\ref{fig:mass_mzr} and \ref{fig:mass_fmr}).

The mass distribution of HGs is affected by the delay time distribution. In our models, different values of the common-envelope efficiency $\alpha$ result in different distributions of the delay time (Appendix~\ref{sec:delay}), with larger values of $\alpha$ being associated with longer delay times.  
Figures \ref{fig:mass_mzr} and \ref{fig:mass_fmr} show that the contribution of high-mass galaxies increases with $\alpha$. In fact, with $\alpha=5$ (longer delay times), the FG has more time to merge with other galaxies and form a more massive HG.  On the other hand, for $\alpha = 1$ (shorter delay times), a large fraction of BBHs are hosted in low-mass galaxies.

In the high-redshift Universe ($z=4$)  both the FGs and HGs of BCOs have a higher SFR than in the local Universe (Figures~\ref{fig:sfr_mzr} and \ref{fig:sfr_fmr}). 
Different values of $\alpha$ have a mild impact on  the shape of the SFR distribution of HGs. On the other hand, the FMR favours  HGs with lower SFR with respect to the MZR in the case of BBHs at $z \ge{}4$. 

The FGs of BBHs and BHNSs tend to have a lower metallicity than their HGs, in the case of both MZR and FMR (Figures \ref{fig:met_mzr} and \ref{fig:met_fmr}). However, the metallicity distributions of both  FGs and HGs strongly depend on the choice of the metallicity relation. If we assume the MZR, the HGs and especially the FGs extend to lower metallicity at high redshift than in the local Universe. In contrast, the FMR predicts relatively high  metallicities for both FGs and HGs and no significant trend with redshift. 

We found a strong correlation between the BNS merger rate per galaxy ($n_{\rm{GW}}$) and the stellar mass of the HG (Figure \ref{fig:mr_z_mass}). 
This correlation is less tight for BBHs and BHNSs, especially if we assume the FMR (Table~\ref{tab:ngw_fits}). 

Passive galaxies can host compact objects mergers (Figure \ref{fig:perc_pass}). Their contribution increases as approaching the local Universe, regardless of the adopted passive galaxy definition. However, the percentage of mergers hosted in passive galaxies  crucially depends on this definition. If  all the galaxies with sSFR $<10^{-10}$ yr$^{-1}$ are considered passive galaxies, we find that $\sim{50-60}$\%, $\sim{45-70}$\% and $\sim{40-75}$\% of all BNS, BHNS and BBH mergers in the local Universe are associated with a passive galaxy, respectively. In contrast, if we define passive galaxies as those galaxies with SFR at least 1 dex below the star-forming main sequence, only $\sim{5-10}$\%, $\sim{15-25}$\% and $\sim{15-35}$\% of all BNS, BHNS and BBH mergers in the local Universe are associated with a passive galaxy, respectively. Overall, BCOs have more chances to be hosted in passive galaxies if their delay time distribution is longer. 


\section*{Acknowledgements}
We thank Martyna Chruslinska, Irina Dvorkin, Stanislav Babak, Leslie K. Hunt, Filippo Mannucci, Giulia Rodighiero, Paolo Cassata and Francesco Sinigaglia for the insightful discussions. MM and FS acknowledge financial support from the European Research Council for the ERC Consolidator grant DEMOBLACK, under contract no. 770017.  MCA acknowledges financial support from 
the  Seal of Excellence @UNIPD 2020 programme under the ACROGAL project. FS thanks the  Astroparticule et Cosmologie Laboratoire (APC) and  the Institute d'Astrophysique de Paris (IAP) for hospitality during the preparation of this manuscript.


\section*{Data Availability}
The data underlying this article will be shared on reasonable request to the corresponding authors. The latest public version of {\sc mobse} can be downloaded from \href{https://gitlab.com/micmap/mobse_open}{this repository}. \cosmorate{} is publicly available on GitLab at \href{https://gitlab.com/Filippo.santoliquido/cosmo_rate_public}{this link}.






\bibliographystyle{mnras}
\bibliography{santoliquido} 

\appendix

\section{\cosmorate{}}
\label{sec:cosmorate}

\cosmorate{} 
estimates the merger rate density based on the average SFR density and metallicity evolution of the Universe, i.e. without considering individual galaxies  \citep{santoliquido2020,santoliquido2021}. 
\cosmorate{} interfaces catalogues of 
simulated BCOs with an observation-based  metallicity-dependent  SFR density evolution of the Universe SFRD$(z,Z)$. 
Namely, \cosmorate{} estimates the merger rate density as
\begin{equation}
\label{eq:mrd}
    \mathcal{R}(z) = \int_{z_{{\rm{max}}}}^{z}\left[\int_{Z_{{\rm{min}}}}^{Z_{{\rm{max}}}} {\rm{SFRD}}(z',Z)\,{} 
    \mathcal{F}(z',z,Z) \,{}{\rm{d}}Z\right]\,{} \frac{{{\rm d}t(z')}}{{\rm{d}}z'}\,{}{\rm{d}}z',
\end{equation}
where 
\begin{equation}
\frac{{\rm{d}}t(z')}{{\rm{d}}z'} = [H_{0}\,{}(1+z')]^{-1}\,{}[(1+z')^3\Omega_{M}+ \Omega_\Lambda]^{-1/2}.
\end{equation}
In the above equation, $H_0$ is the Hubble constant, $\Omega_M$ and $\Omega_\Lambda$ are the matter and energy density, respectively. We adopt the values in \cite{Planck2016}. 
The term $\mathcal{F}(z',z,Z)$ is given by:
\begin{equation}
\mathcal{F}(z',z,Z) = \frac{1}{\mathcal{M}_{{\rm{TOT}}}(Z)}\frac{{\rm{d}}\mathcal{N}(z',z, Z)}{{\rm{d}}t(z)},
\end{equation}
where $\mathcal{M}_{{\rm{TOT}}}(Z)$ is the total simulated initial stellar mass, and  ${{\rm{d}}\mathcal{N}(z',z, Z)/{\rm{d}}}t(z)$ is the rate of BCOs forming from stars with initial metallicity $Z$ at redshift $z'$ and merging at $z$, extracted from our population-synthesis catalogues. 
In \cosmorate{},    ${\rm{SFRD}}(z,Z)$  is given by 
\begin{equation}
{\rm{SFRD}}(z',Z) = \psi(z')\,{}p(z',Z),
\end{equation}
where $\psi(z')$ is the cosmic SFR density at formation redshift $z'$ from \cite{madau2017}, and $p(z',Z)$ is the log-normal distribution of metallicities $Z$ at fixed formation redshift $z'$,
with average $\mu(z')$ and spread $\sigma_{Z}$. Here, we 
consider two definitions of $\mu(z)$. Equation 4 of  \cite{santoliquido2020}, hereafter model S20:
\begin{equation}
    \mu(z) = \log_{10}{a} + b\,{} z, 
\end{equation}
where $a = 1.04 \pm 0.14$ and $b = -0.24 \pm 0.14$
; and Equation 6 of \cite{madau2017}, hereafter MF17, that is:
\begin{equation}
\mu(z) = 0.153 - 0.074\,{} z^
{1.34} - \frac{{\ln{(10)}}\,{}\sigma_{Z}^2}{2}
\end{equation}
For these models, we  considered two different values of the dispersion: $\sigma_Z = 0.2$ for S20 and  $\sigma_Z = 0.3$ for MF17. In Figure~\ref{fig:mrd} we also explore $\sigma_Z = 0.4,$ 0.6 and 0.7 for MF17. 

\section{Delay time distributions and merger efficiency}
\label{sec:delay}

\begin{figure}
    \centering
    \includegraphics[width = 0.45\textwidth]{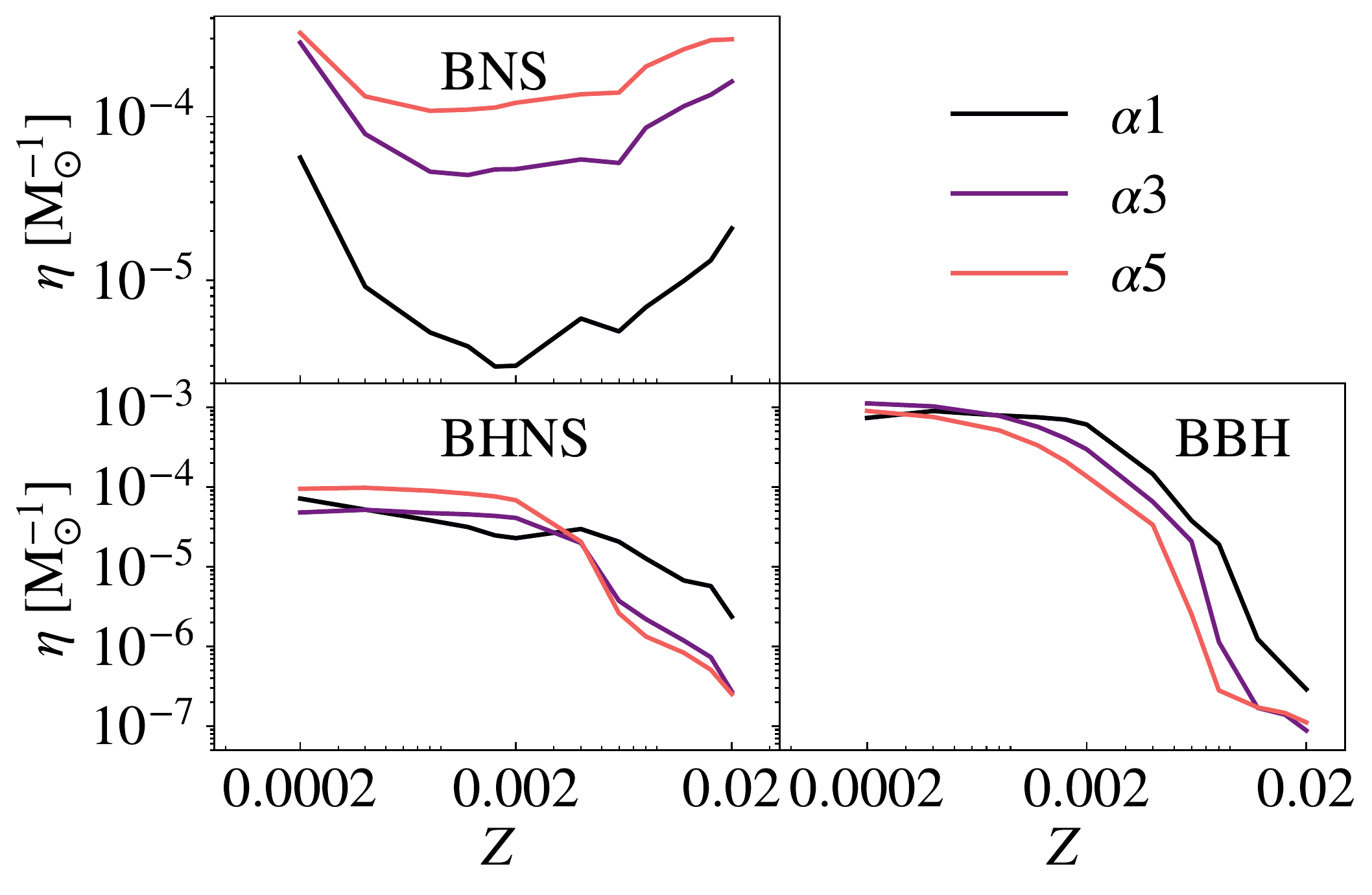}
    \caption{Merger efficiency $\eta$ as a function of progenitor metallicity for models $\alpha=1,~3$ and $5$.}
    \label{fig:eta}
\end{figure}

\begin{figure*}
    \centering
    \includegraphics[width = 0.85\textwidth]{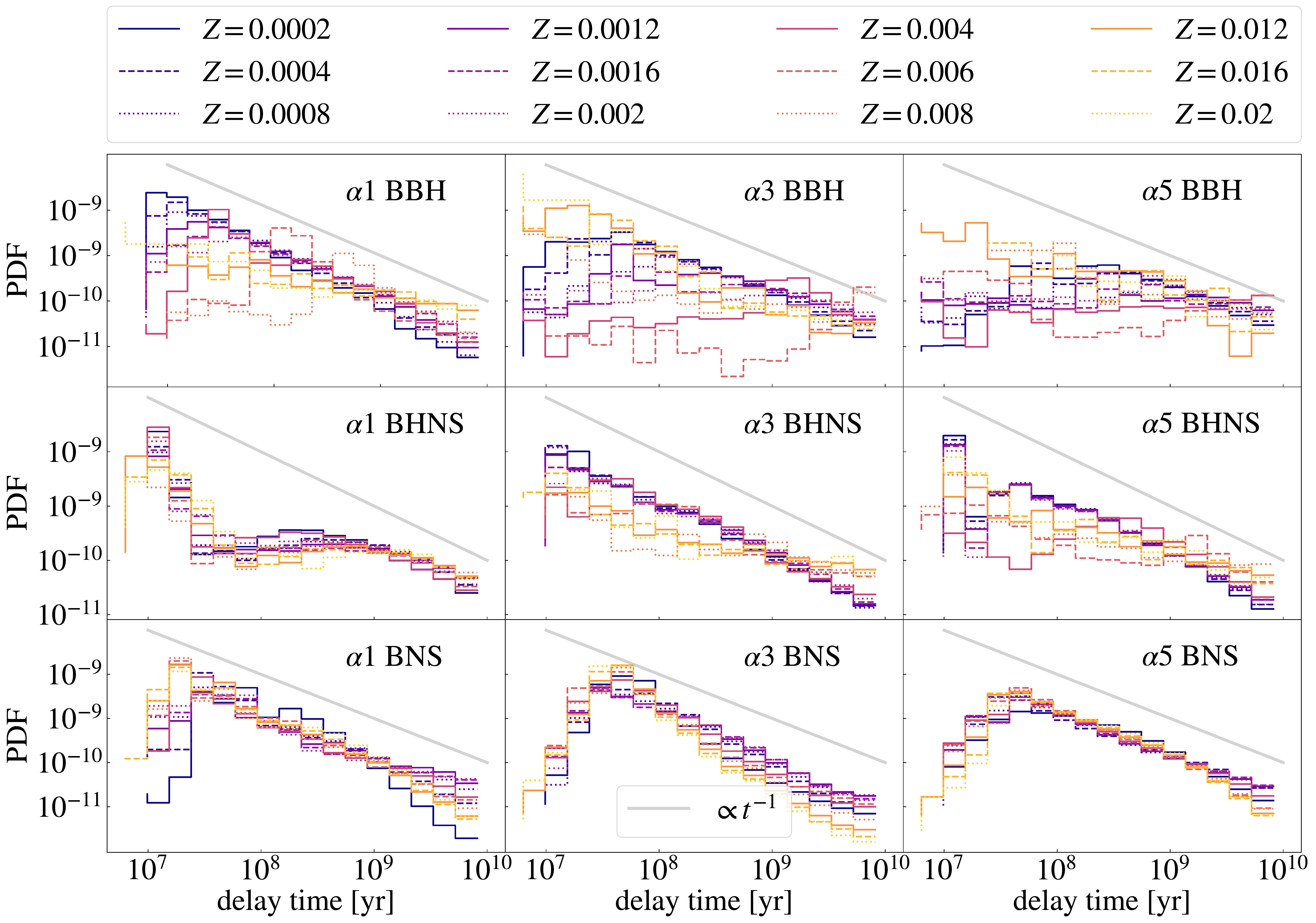}
    \caption{From left to right: delay time distributions for $\alpha=1,~3$ and $5$ for BBHs (upper panels), BHNSs (middle) and BNSs (lower panels). Each color refers to a different progenitor's metallicity. The grey line is  $\propto t_{\rm del}^{-1}$ with arbitrary normalisation. }
    \label{fig:delaytime}
\end{figure*}


The merger efficiency and delay time distribution of BBHs strongly depend on metallicity and $\alpha{}$.
Figure \ref{fig:eta} show the merger efficiency $\eta (Z)$, defined as:
\begin{equation}
\label{eq:eta}
    \eta (Z) = f_{\rm bin}f_{\rm IMF}\,{}  \frac{\mathcal{N}_{\text{TOT}}(Z)}{M_\ast{}(Z)},
\end{equation}
where $f_{\rm{bin}} = 0.5$ is the binary fraction \citep{sana2012}, and $f_{\rm{IMF}}$ is a correction factor taking into account that we simulated only stars with mass $m>5$ M$_\odot$ with {\sc{mobse}}. Thus, assuming a Kroupa IMF \citep{kroupa2001}, $f_{\rm{IMF}} = 0.285$.  

Figure~\ref{fig:eta} shows that the merger efficiency decreases by four orders of magnitude if we vary $Z = 0.0002$ to $Z = 0.02$. This strong dependence of $\eta$ on $Z$ has been already widely described in the literature and is common to very different population-synthesis codes (e.g. \citealt{dominik2012,giacobbo2018a,klencki2018,neijssel2019,spera2019}). It is a consequence of stellar winds:  metal-poor stars retain larger masses during their lives and grow larger radii than metal-rich ones. This enhances their chances to undergo mass transfer with companion stars and produce tight BBHs.
 
 The merger efficiency of BNSs varies much less with $Z$ but is heavily affected by $\alpha$. The parameter $\alpha$ also shapes the distribution of delay times. 
 Figure \ref{fig:delaytime} shows that 
 if $\alpha=1$ 
 the delay time distribution peaks at shorter delay times than for larger values of $\alpha$. Smaller values of $\alpha$  correspond to shorter delay times, because a small value of $\alpha{}$ implies a more effective shrinking of the progenitor binary during common envelope. Hence, binary compact objects tend to form with shorter orbital separation if we assume a lower value of $\alpha$. 
Also, Figure \ref{fig:delaytime} shows that ${\rm d}N/{\rm d}t_{\rm del}\propto{}t_{\rm del}^{-1}$ for long delay times. 

\section{Impact of the comoving volume on the merger rate}
\label{sec:1gpc_vol}

\begin{figure}
    \centering
    \includegraphics[width = 0.45\textwidth]{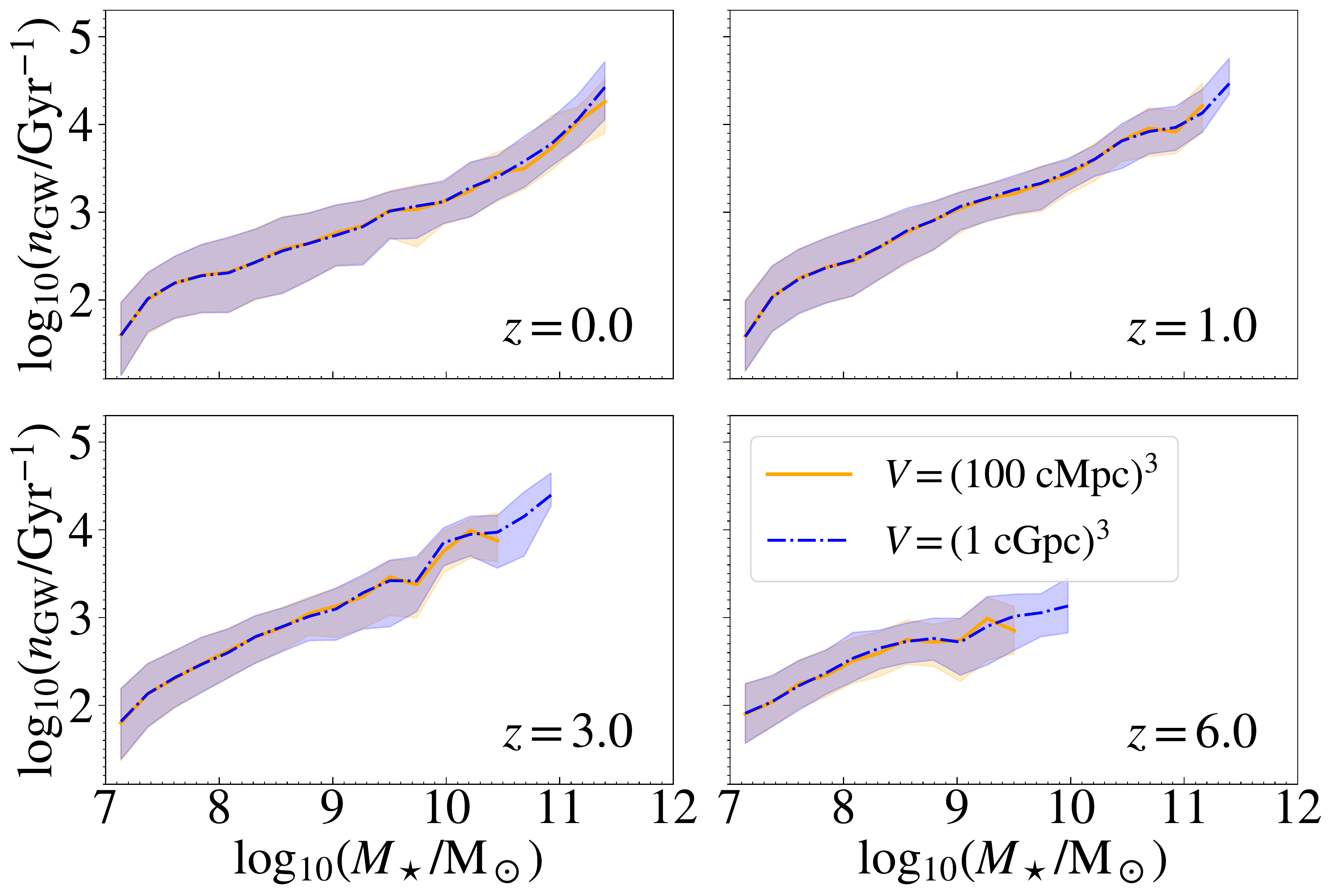}
    \caption{BBH merger rate per galaxy ($n_{\rm{GW}}$) as function of stellar mass ($M_*$) for two different values of the comoving volume: $V = (100 ~{\rm{cMpc}})^3$ and $V = (1 ~{\rm{cGpc}})^3$. We assume $\alpha=5$, FMR and B18.}
    \label{fig:1gpc_vol}
\end{figure}

We assumed a comoving volume $V = (100~ {\rm{cMpc}})^3$ to obtain the results presented in the main text. This choice is a compromise between the need to sample a large volume and the computational cost: it takes $\sim 15$ and $\sim 2.5 \times 10^{2}$ CPU hours to simulate a volume $V = (100~ {\rm{cMpc}})^3$ and $V = (1~ {\rm{cGpc}})^3$, respectively.
Figure \ref{fig:1gpc_vol} compares the case in which we assume $V=(100~{\rm cMpc})^3$ and $(1~ {\rm{cGpc}})^3$ for one test case (with $\alpha{}=5$, FMR and B18). We find statistically no differences between the two volumes at redshift $z=0$, while at higher redshift considering a large volume allows us to include more massive galaxies. Overall, we find no differences in the mass range we considered in this study, and our main results can be easily extrapolated to larger galaxy masses.

\section{Impact of the minimum galaxy mass on the merger rate}
\label{sec:min_mass}

\begin{figure}
    \centering
    \includegraphics[width = 0.40\textwidth]{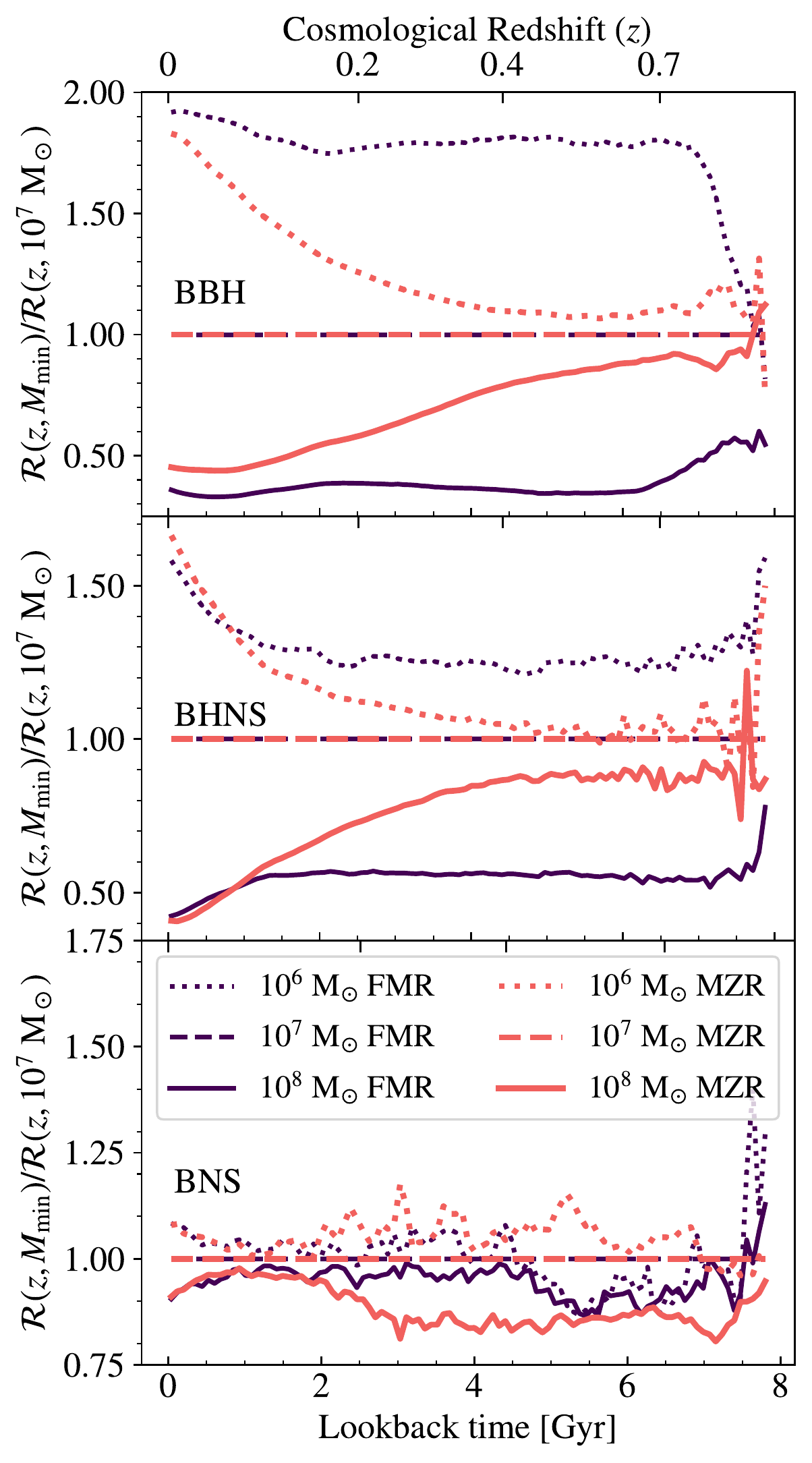}
    \caption{Impact of the choice of the minimum galaxy stellar mass ($M_{\rm min}$) on the merger rate density of BBHs (upper panel), BHNSs (middle) and BNSs (lower). Each line is the ratio of the merger rate density $\mathcal{R}(z)$ we estimate assuming $M_{\rm min}=10^6$ (dotted line), 10$^7$ (dashed line), and 10$^{8}$ M$_\odot$ (solid line) to the merger rate density obtained for our fiducial value $M_{\rm min}=10^7$~M$_\odot$. Blue lines: FMR. Pink lines: MZR. We show the case with $\alpha=5$ and for S14.} 
    \label{fig:mrd_mimmass}
\end{figure}

In our main text, we adopted a value $M_{\rm min}=10^7$ M$_\odot$ for the minimum galaxy stellar mass. In the local Universe, we see galaxies with mass lower than $10^7$ M$_\odot$, but it is not clear if we can extrapolate the main scaling relations (GSMF, SFR, MZR and FMR) down to such small masses. Actually, there is a mild evidence that the lowest mass star forming galaxies deviate from the main observational scaling relations \citep[e.g.,][]{hunt2012}. 

Figure \ref{fig:mrd_mimmass} shows the impact of two other choices of the minimum galaxy stellar masses, namely $10^6$, and $10^8$ M$_\odot$. In particular, we show the ratio of the merger rate density we obtain by varying the minimum galaxy stellar mass to the merger rate density we obtain for the fiducial value $ M_{\rm min}=10^7$ M$_\odot$. 

A lower (higher) value of $M_{\rm min}$ implies a higher (lower) merger rate density for both BBHs and BHNSs, because of their dependence on the metallicity of the progenitor stars. We find a maximum difference of a factor of two between the BBH merger rate density with $M_{\rm min}=10^6$ M$_\odot$ and  $10^7$ M$_\odot$. This difference is nearly constant across redshift for the FMR, while it becomes smaller at high redshift for the MZR. In contrast,  BNSs are  not affected by the choice of $M_{\rm min}$.

\section{Impact of the solar metallicity on the merger rate density}
\label{sec:met_sol}

\begin{table}
    \centering
    \begin{tabular}{l|l|c|c}
    \toprule
         Reference &  & $Z_\odot$ & $12 + \log_{10}({\rm{O/H}})_\odot$\\
         \hline
         \cite{Asplund2009} & A09 & 0.0134 & 8.69 \\
        \cite{caffau2011} & C11 & 0.0153 & 8.76\\
         \cite{anders1989} & AG89 & 0.017 & 8.83 \\
        \cite{villante2014} & V14 & 0.019 & 8.85\\
         \cite{grevesse1998} & GS98 & 0.0201 & 8.93 \\
    \bottomrule
    \end{tabular}
    \caption{
    {Solar metallicity measures 
    from different authors. }}
    \label{tab:sol_met}
\end{table}

\begin{figure}
    \centering
    \includegraphics[width = 
    0.40\textwidth]{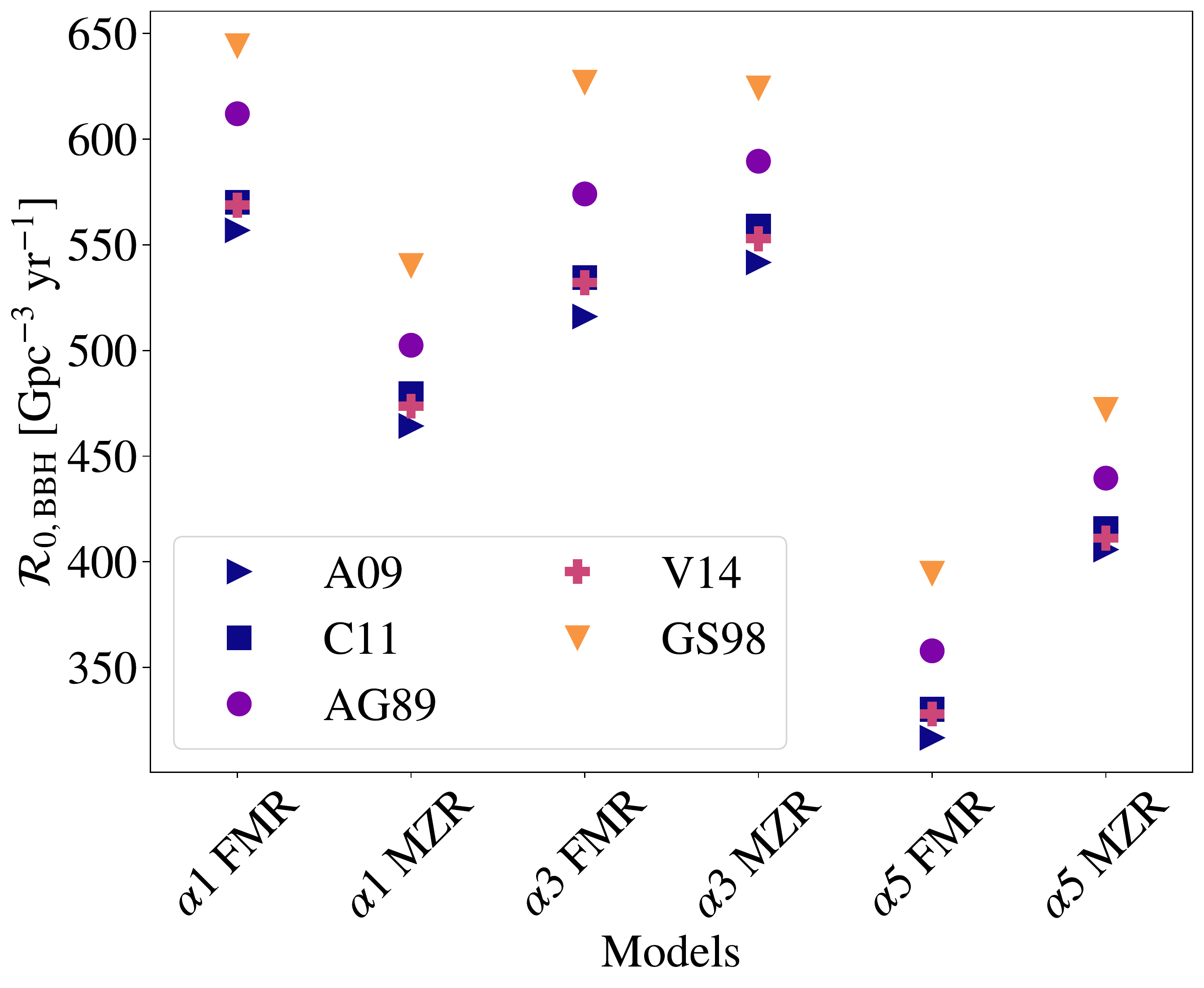}
    \caption{BBH merger rate density in the local Universe evaluated with different definitions of solar metallicity: A09 refers to \protect\cite{Asplund2009}, V14 to  \protect\cite{villante2014}, C11 to \protect\cite{caffau2011}, GS98 to \protect\cite{grevesse1998} and AG89 to \protect\cite{anders1989}. See Table \ref{tab:sol_met} for more details. We show the results for $\alpha=1,$ 3, 5, and for both the MZR and FMR, assuming the main sequence definition from S14.}
    \label{fig:mrd_zsun}
\end{figure}

Both the MZR and FMR are expressed in terms of the relative abundance of oxygen and hydrogen $12 + \log_{10}(\rm{O/H})$. In order to convert this quantity to the  mass fraction of all elements heavier than helium $Z$, we assumed, as commonly done in literature \citep[e.g.][]{maiolino2019}, that the latter scales linearly with the measured $12 + \log_{10}(\rm{O/H})$. In other words, we assumed that $Z$  maintains the solar abundance ratio, as expressed by the following equation:
\begin{equation}
    \log_{10}{Z} = \log_{10}Z_\odot + \log_{10}({\rm{O/H}}) - \log_{10}({\rm{O/H}})_\odot.
\end{equation}
Figure \ref{fig:mrd_zsun} shows the impact of different definitions of the solar metallicity (Table \ref{tab:sol_met}) on the merger rate density of BBHs, which are the most affected by metallicity variations among BCOs. 
We show only the local merger rate density since changing the definition of the solar metallicity only affects the normalisation of the merger rate density.


\bsp	
\label{lastpage}
\end{document}